\documentclass[a4paper,12pt]{iopart}

\usepackage{amssymb}
\usepackage{amsthm}
\usepackage{amsfonts}
\usepackage{url}
\usepackage{hyperref}
\usepackage{breakurl}
\usepackage{rotating}
\usepackage{adjustbox}
\usepackage{longtable}
\usepackage{topcapt}
\usepackage{pdflscape}
\usepackage{microtype}
\usepackage{caption}

\theoremstyle{plain}
\renewcommand{\thetable}{\arabic{table}}

\newcommand{\seqnum}[1]{\href{http://oeis.org/#1}{\underline{#1}}}

\begin{document}

\title{Self-avoiding walks contained within a square}

\author{Anthony J Guttmann}
\address{School of Mathematics and Statistics,
The University of Melbourne,
Vic. 3010, Australia}
\ead{guttmann@unimelb.edu.au}

\author{Iwan Jensen}
\address{College of Science and Engineering, Flinders University at Tonsley,
GPO Box 2100, Adelaide, SA 5001, Australia}
\ead{iwan.jensen@gmail.com}

\author{Aleksander L Owczarek}
\address{School of Mathematics and Statistics,
The University of Melbourne,
Vic. 3010, Australia}
\ead{owczarek@unimelb.edu.au}

\begin{abstract}
We have studied self-avoiding walks contained within an $L \times L$ square whose  end-points can lie anywhere within, or on, the boundaries of the square. We prove that such walks behave, asymptotically, as walks crossing a square (WCAS), being those walks whose end-points lie at the south-east and north-west corners of the square.

We provide numerical data, enumerating all such walks, and analyse the sequence of coefficients in order to estimate the asymptotic behaviour. 
We also studied a subset of these walks, those that must contain at least one edge on all four boundaries of the square. We provide compelling evidence that these two classes of walks grow identically.

From our analysis we conjecture that the number of such walks $C_L$, for both problems, behaves as $$ C_L \sim \lambda^{L^2+bL+c}\cdot L^g,$$ where \cite{GJ22} $\lambda=  1.7445498 \pm 0.0000012,$ $b=-0.04354 \pm 0.0005,$ $c=-1.35 \pm 0.45,$ and $g=3.9 \pm 0.1.$

Finally, we also studied the equivalent problem for self-avoiding polygons, also known as cycles in
a square grid. The asymptotic behaviour of cycles has the same form as walks, but with different values of the parameters $c$, and $g$. Our numerical analysis shows that $\lambda$ and $b$ have the same values as for WCAS and that $c=1.776 \pm 0.002$ while $g=-0.500\pm 0.005$ and hence probably equals $-\frac12$.
\end{abstract}

\section{Introduction}
\label{introduction}
The behaviour of self-avoiding walks (SAWs) crossing a square has a long and interesting history. The canonical problem (of which there are several variants) considers SAWs in an $L \times L$ domain with origin at $(0,0)$ and end-point at $(L,L)$ and confined to lie entirely within the the square domain (including its boundaries). 
Some history can be found in \cite{GJ22} and in the On-line Encyclopaedia of Integer Sequences \cite{OEIS}, \seqnum{A007764}.
It is known that the number of such walks grows as $\lambda_S^{L^2},$ where the best estimate of $\lambda_S$ is $1.7445498 \pm 0.0000012$ \cite{GJ22}. We will refer to these walks as ``walks crossing a square" or WCAS for short.

Recently a related problem was studied by Bradly and Owczarek in \cite{BO21}, notably the problem of SAWs in an $L \times L$ square, but with no constraints placed on the origin or end point. That is to say, they are SAWs that can lie anywhere in or on the boundary of the square. 
In \cite{BO21} many properties of these more numerous walks were investigated, and the possibility that their growth might be similar to that of WCAS was suggested.
Here we provide a proof of this result, and extensive numerical data that not only (unnecessarily) supports this conclusion, but allows us to estimate the sub-dominant behaviour, and so give an expression for the asymptotic behaviour, beyond leading order.

\section{Proof of key result} \label{proof}
Consider SAWs on the square lattice which are wholly contained in a square of side $L$ bonds (with $(L+1)$ vertices) so that 
all sites of the walk lie within or on the boundary of the square. The length of the walks $n$ being the numbers of steps is not fixed. 

\begin{figure}%
	\centering
	\includegraphics[width=0.3\columnwidth]{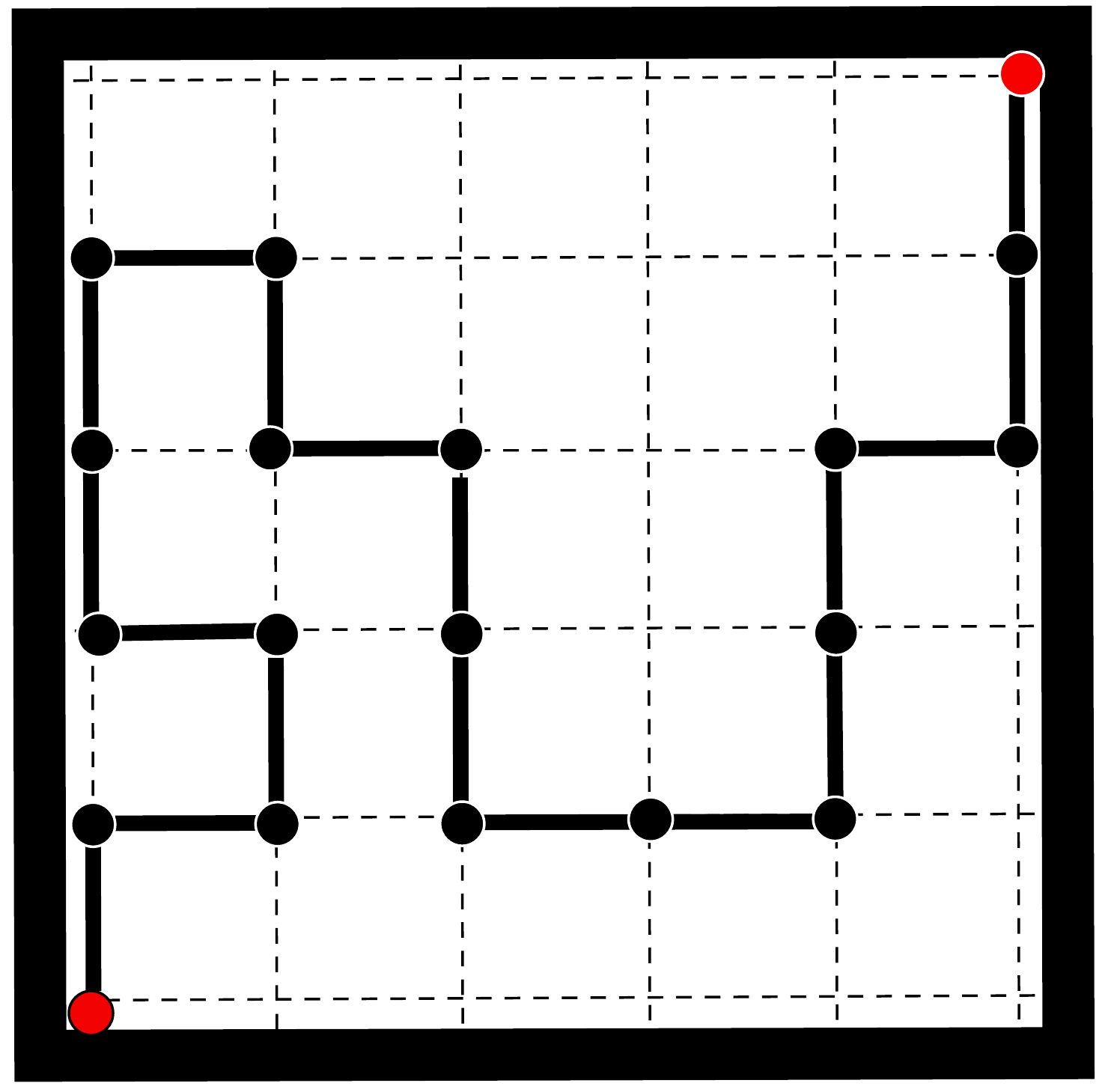}%
	\includegraphics[width=0.3\columnwidth]{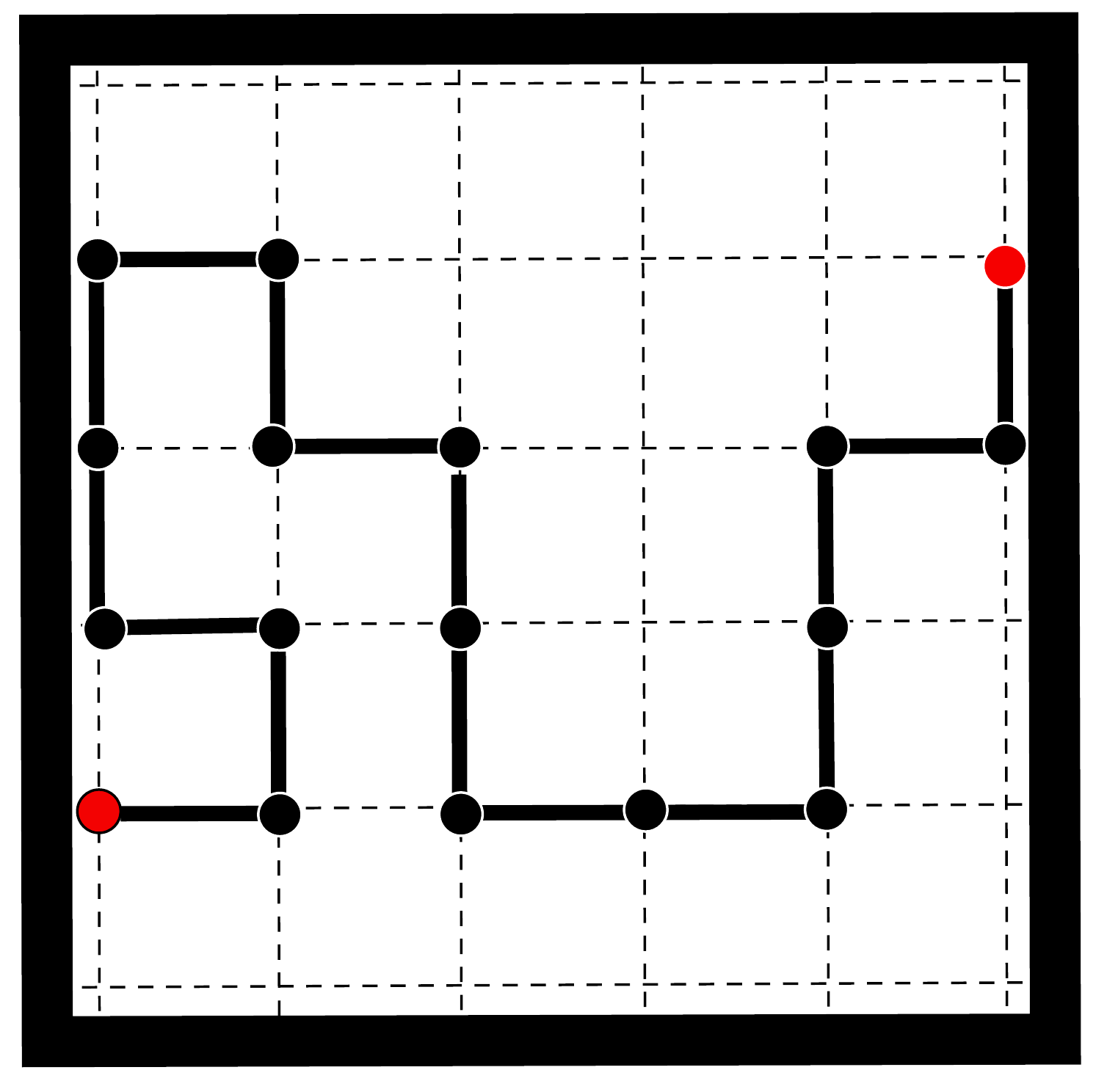}%
	\includegraphics[width=0.3\columnwidth]{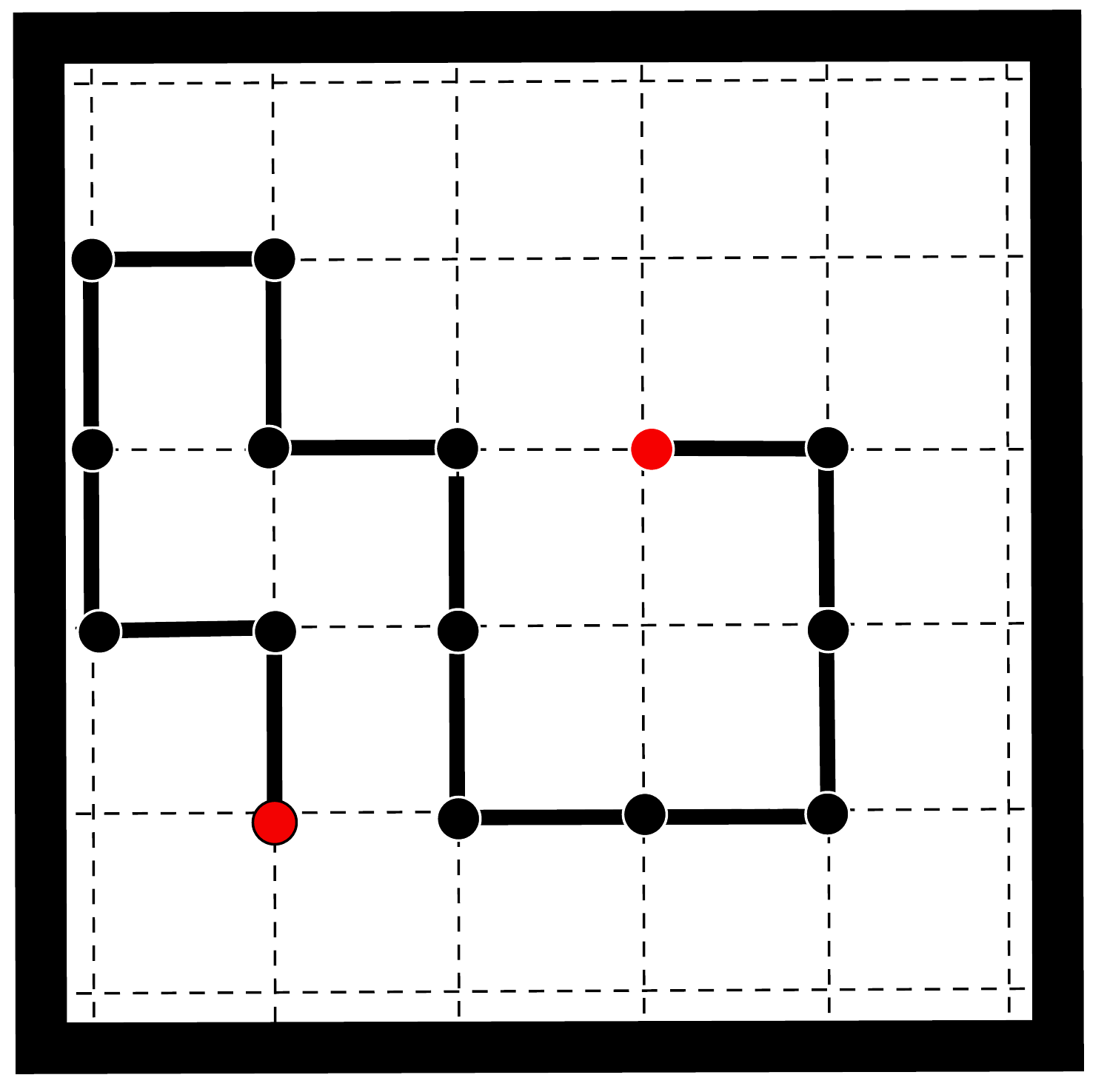}%
	\caption{From left to right are examples of walks whose endpoints lie at opposing corners of the square, whose endpoints lie on opposing sides of the square, and whose endpoints lie anywhere inside the square (or on the boundary).}%
	\label{fig:Configurations}%
\end{figure}

We consider three conditions of the endpoints of the SAW (See \Fref{fig:Configurations}) and the numbers of such walks as
\begin{enumerate}
\item walks whose endpoints lie at opposing corners of the square counted as $R_L$;
\item walks whose endpoints lie on opposing sides of the square  counted as $S_L$;
\item walks whose endpoints lie anywhere inside the square (or on the boundary) counted as  $A_L$.
\end{enumerate}

Let us further define the number of walks  whose endpoints lie anywhere inside a square of side $\ell$ that span the square in at least one direction: in this way they are walks whose minimal bounding box is of side $\ell$ and these have cardinality $M_\ell$ with the convention $M_0=1$.

Denote by $\widehat{M}_L$ the union of the sets counted by $M_\ell$ for $0\leq \ell \leq L.$ Hence
\begin{equation}
\widehat{M}_L = \sum_{\ell =0}^L M_\ell
\end{equation}
and 
\begin{equation}
R_L \leq S_L \leq M_L \leq \widehat{M}_L \leq A_L\;.
\end{equation}
Furthermore since walks counted by $\widehat{M}_L$  can be translated no more than $(L+1)^2$ times to give a configuration counted by $A_L$ we have
\begin{equation}
\widehat{M}_L \leq A_L \leq (L+1)^2 \widehat{M}_L\;.
\label{Abound}
\end{equation}
It is known \cite{AH78,GW90} that 
\begin{equation}
\lambda =  \lim_{L \to \infty}  R_L^{1/L^2}\;.
\end{equation} exists  so that 
\begin{equation}
R_L = \lambda^{L^2 +o(L^2)} \mbox{ as } L \to \infty \; .
\end{equation}
It is not difficult to deduce that
\begin{equation}
S_L = \lambda^{L^2 +o(L^2)} \mbox{ as } L \to \infty\; .
\end{equation}
In this work we are interested in the limit
\begin{equation}
\lambda_A =  \lim_{L \to \infty}  A_L^{1/L^2}\;.
\label{lambdaA}
\end{equation}
However, we begin with 
\begin{equation}
\lambda_{\widehat{M}} = \lim_{L \to \infty}  \widehat{M}_L^{1/L^2}
\label{lambdaMhat}
\end{equation}
Now (\ref{Abound}) implies that if the limit (\ref{lambdaMhat}) exists the limit (\ref{lambdaA}) exists
and 
\begin{equation}
\lambda_A= \lambda_{\widehat{M}} \;.
\end{equation}

First consider the set of walks counted by $M_\ell$. On two sides of the square, at least, some vertices of the walk lie on the opposed boundaries. Therefore either the end point lies on the boundary or there is a step of the walk on the boundary with or without an endpoint on the boundary. Choose the left (bottom) and right (top) sides of the box as those that the walk touches. Extend the walk by choosing the first boundary site of the walk moving clockwise from the bottom left (bottom right) corner. If this is an endpoint, just add a horizontal (vertical) step outside the box,  and if not an endpoint itself it must be attached to a boundary step so replace the boundary step by three steps that surround the plaquette outside the bounding box. We always extend the left hand (bottom) side of the configuration. This gives a unique configuration in the set counted by $M_{\ell +1}$. Hence
\begin{equation}
M_{\ell+1} \geq M_\ell
\end{equation}
so that 
\begin{equation}
M_\ell \leq M_L \mbox{ for all } \ell < L
\end{equation}
and 
 \begin{equation}
M_L < \widehat{M}_L = \sum_{\ell =0}^L M_\ell  <  (L+1) M_L \mbox{ for  } L\geq 1
\end{equation}
which implies that if the limit 
\begin{equation}
\lambda_M= \lim_{L \to \infty}  M_L^{1/L^2}
\end{equation}
exists, then the limit (\ref{lambdaA}) also exists and $\lambda_M=\lambda_{\widehat{M}}$.

We now show that $\lambda_M$ exists and that
\begin{equation}
\lambda_M=\lambda
\label{our_result}
\end{equation}
so that
\begin{equation}
\lambda_A = \lambda_{\widehat{M}} =\lambda_M=\lambda\;.
\end{equation}

Consider a configuration in the set counted by $M_L$ that spans left to right. If a configuration only spans the square bottom to top all the following goes through by making the appropriate substitutions of bottom for left and top for right. Let the bottom left hand corner of the square be at the origin of the square lattice, and the top right corner is at $(L,L)$. Let the two end points be at sites $(x_1,y_1)$ and $(x_2,y_2)$ and $0\leq  x_1 \leq x_2\leq L $. If the endpoints occur in the same column, choose the uppermost one to be the `rightmost'. The sum of the distance of the leftmost end point  $(x_1,y_1)$ to the left boundary at $x=0$ plus the distance of the rightmost end point to the right boundary at $x=L$ is
\begin{equation}
x_1+ (L-x_2) = L - (x_2-x_1) \leq L\;.
\end{equation}

Now consider an algorithm whose procedure is:  move the end points to the boundaries of the square in a such a way that the leftmost end point  $(x_1,y_1)$ is moved to the left boundary at $x=0$ and the rightmost end point is moved to the right boundary at $x=L$. Again wlog consider moving the rightmost end point to the right boundary using the following possible moves: The first two come from the backbite algorithm (Mansfield, 1982) and are called the backbite move and the end-attack move. We actually only consider a subset of those moves that definitely move the endpoint to the right. It is not always possible to use these moves to do this, and so we introduce two further moves that either lengthen or shorten to the walk by adding or deleting a horizontal step. 

\begin{figure}%
	\centering
	\includegraphics[width=0.7\columnwidth]{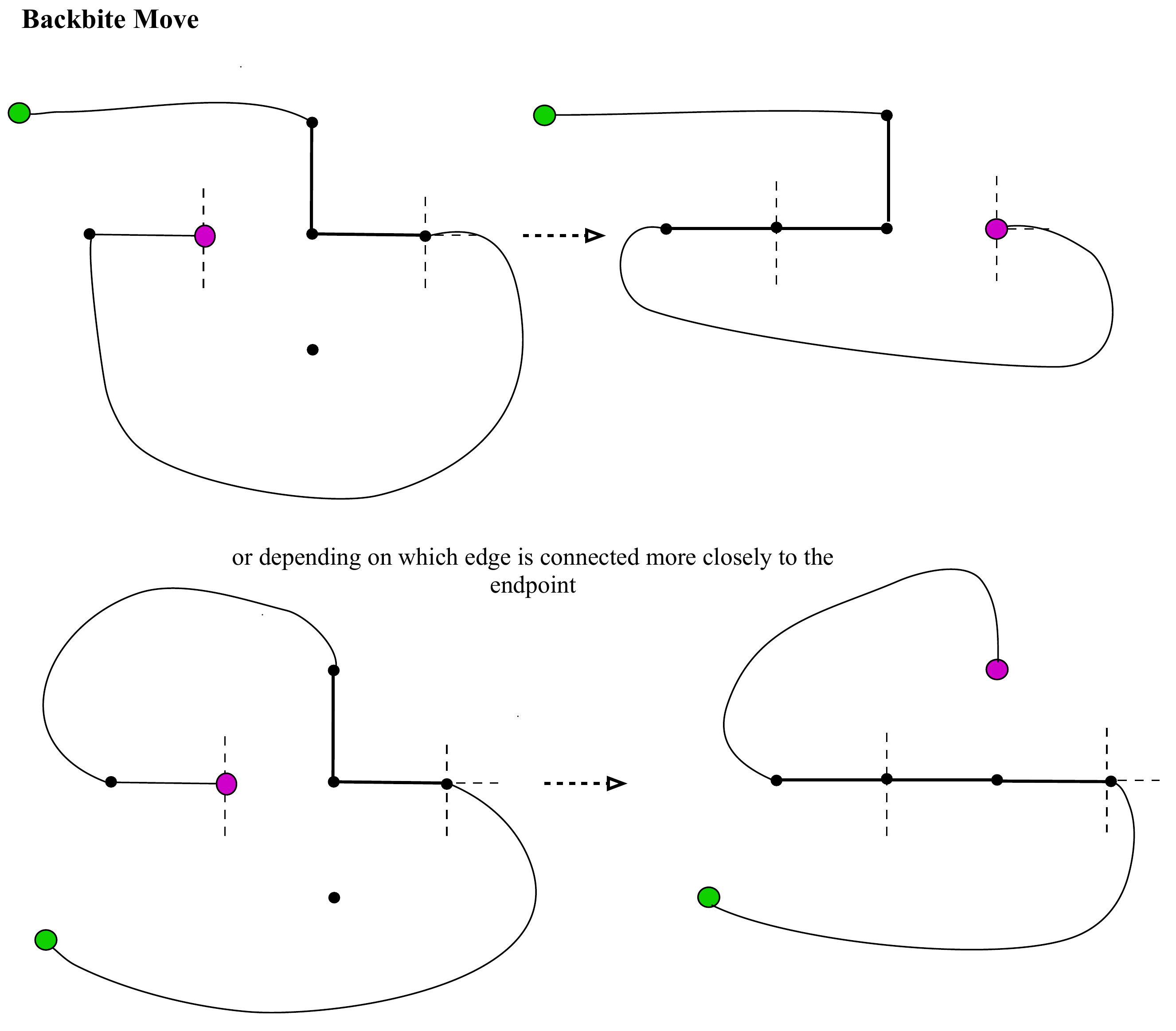}\\%
	\includegraphics[width=0.7\columnwidth]{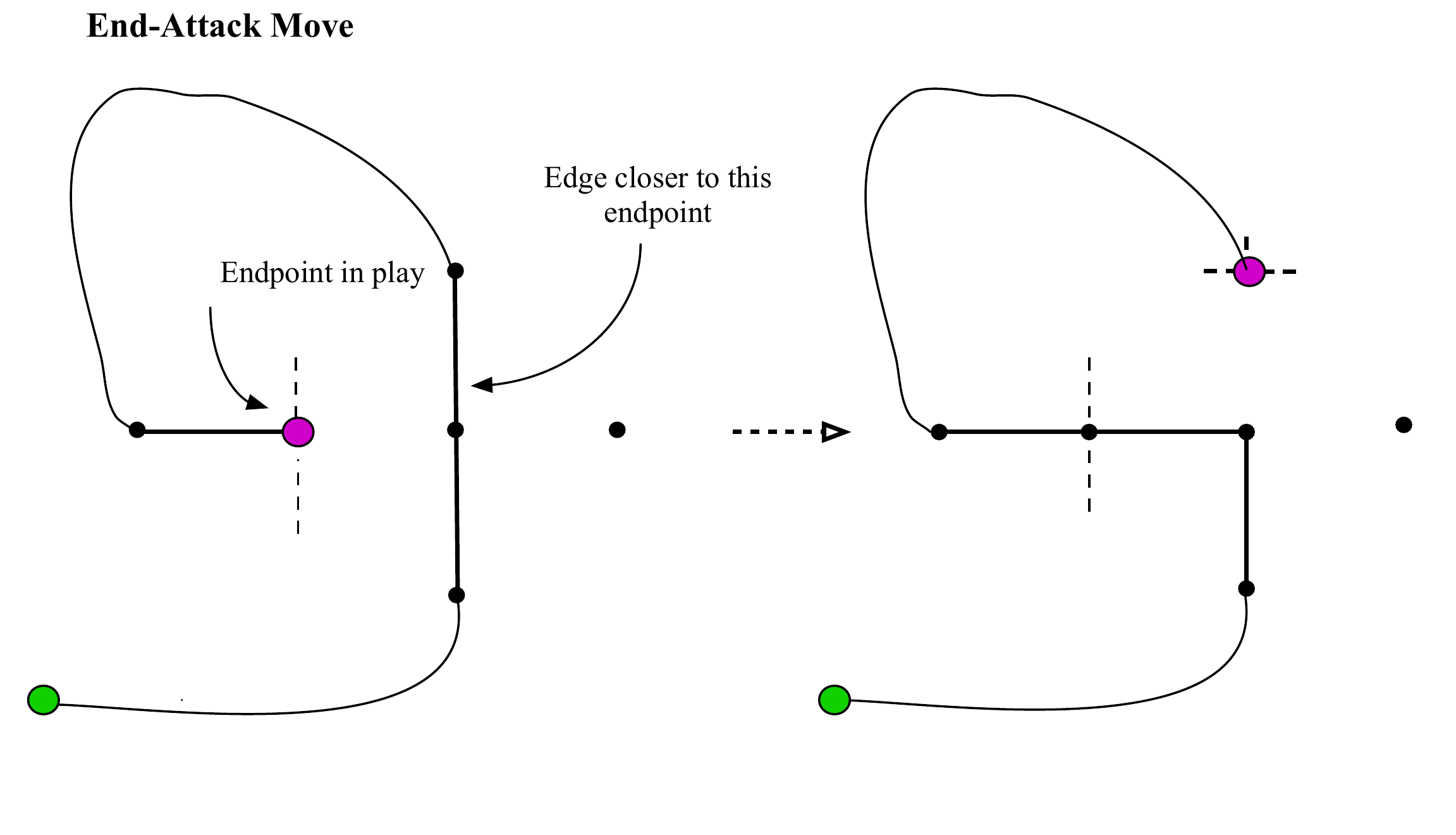}\\%
	\includegraphics[width=0.7\columnwidth]{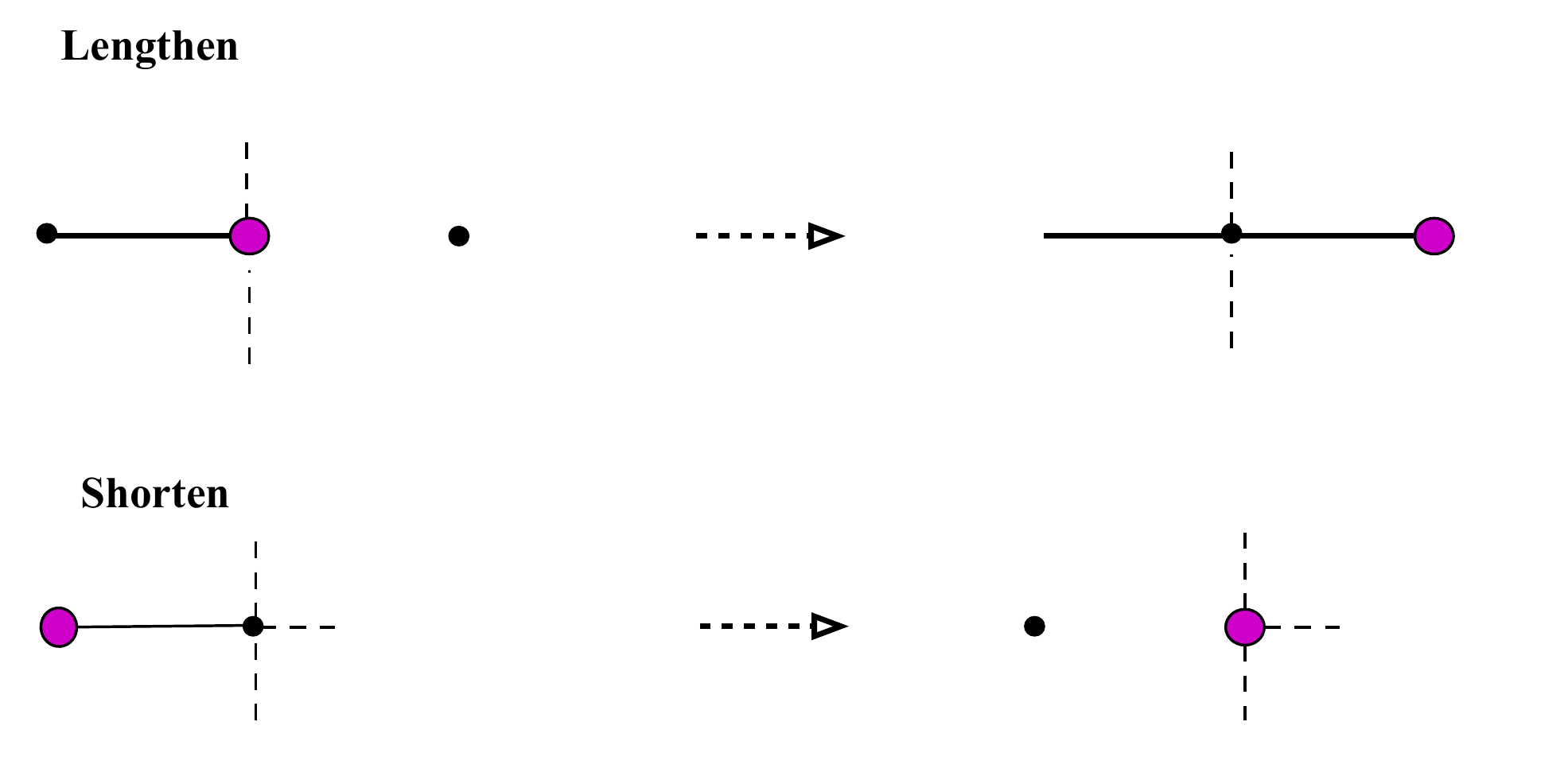}%
	\caption{The possible moves in the algorithm that moves an endpoint to the right: the backbite moves at the top, the end-attack move in the centre and the lengthening and shortening moves at the bottom.}%
	\label{fig:Moves}%
\end{figure}

All four possible moves are shown in \Fref{fig:Moves}. The choice of move is dictated by the local configuration to the immediate right of the endpoint. If the site to the immediate right of the endpoint is unoccupied we lengthen the walk and move the endpoint one lattice edge towards the right boundary. 

If the edge of the lattice to the immediate right of the endpoint is occupied we delete that step, shortening the walk and again move the endpoint one lattice edge towards the right boundary. 

If the site to the immediate right of the endpoint is occupied but the edge to the immediate right is not occupied, then we either use the end-attack or backbite moves as shown in \Fref{fig:Moves}. Given that our endpoint is rightmost, the only two possibilities for the edges incident on the site to the immediate right of the end point are (i) that the two vertical edges of the lattice incident on that site are both occupied or (ii) one vertical edge is occupied and the next horizontal edge is occupied. This implies that we must be able to use one of the moves \Fref{fig:Moves} to move the endpoint to the right. Note that sometimes the backbite move actually moves the endpoint two columns to the right. The number of times we need to apply this procedure is less than the distance to the boundary, and so in moving both endpoints we apply the algorithm at most $L$ times. So every configuration in the set counted by $M_L$ can be mapped by this procedure to a configuration in the set counted by $S_L$.

\begin{figure}[h!]%
	\centering
	\includegraphics[width=0.7\columnwidth]{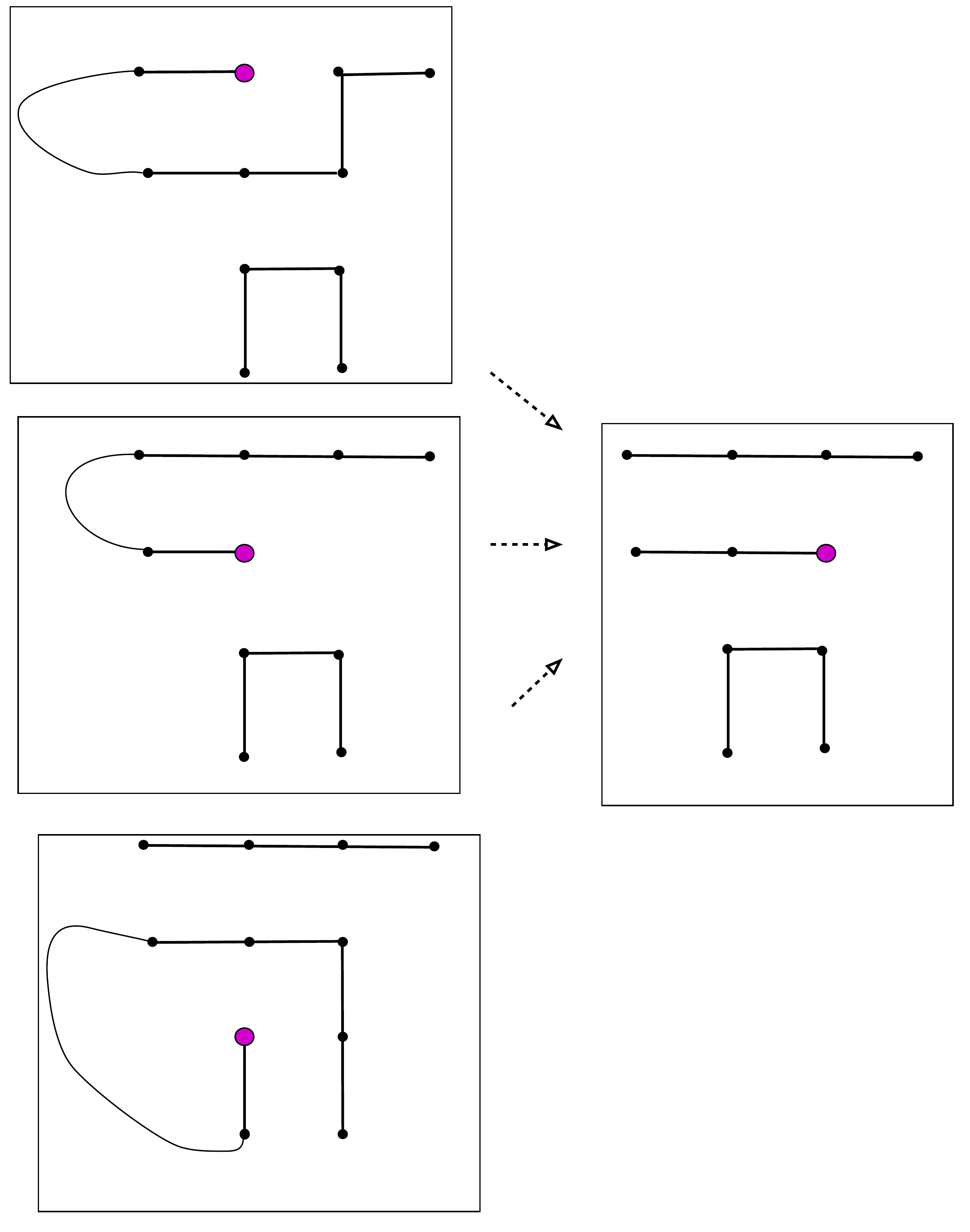}%
	\caption{On the left are the three possible antecedents of the configuration on the right when moving the endpoint to the right.}%
	\label{fig:Antecedents}%
\end{figure}

Now let us consider the multiplicity of configurations counted by $M_L$ that are mapped to any particular configuration in the set counted by $S_L$. By considering the antecedents of the endpoints of an intermediate configuration that eventually maps to a particular configuration in the set counted by $S_L,$ these must have been able to use one of our moves to obtain the current one. To make a particular site an endpoint, either a step has been deleted or one of three were added (or one of two if deletion is possible, noting that only one of the lengthening or shortening moves could have been used). This means that a maximum of three different antecedents may occur. For example see \Fref{fig:Antecedents}. 

Consider the antecedents whose left endpoint reach the boundary after $b$ moves, then the total is a sum over values of $b\leq L$. Hence the total number of antecedents is bounded for $L\geq 2$ by 
\begin{equation}
\sum_{b=0}^L 3^b \sum_{r=0}^{L-b} 3^r = \frac{1}{2}\sum_{b=0}^L  3^{b}( 3^{L- b +1} -1 )<  \frac{1}{2} (L+1)  3^{L+1} \;.
 \end{equation}
Hence we have 
\begin{equation}
S_L \leq M_L \leq \frac{1}{2} (L+1)  3^{L+1} S_L
\end{equation}
and the result (\ref{our_result}) follows by raising each to the power $1/L^2$ and taking the limit $L \rightarrow \infty$. 

\newpage
\section{Numerical analysis} \label{numerical}

The algorithm for counting SAWs contained in a square is a simplified version of the transfer matrix algorithm of \cite{IJ04} used to count SAWs on the square lattice with a step fugacity. 
Instead of counting SAWs with a fugacity we no longer keep track of the number of steps (similar to the algorithms from \cite {BGJ05} used to count walks crossing or spanning a square).
The algorithm counts the number of walks $\widehat{A}_{h,l}$ in a rectangle of size $h\times l$, where the walks touch {\em all} four boundaries of the rectangle. By symmetry clearly $\widehat{A}_{l,h}=\widehat{A}_{h,l}$ so we need only consider the case $l\geq h$.
A rectangle of size $h\times l$ can be placed inside an $L\times L$ square ($L \geq l$) in $(L-l+1)\cdot (L-h+1)$ different positions. Hence

$$A_L = \sum_{l=1}^L(L-l+1)^2 \widehat{A}_{l,l} \;\; + \;\; 2\sum_{l=h+1}^L \sum_{h=0}^L (L-l+1) (L-h+1) \widehat{A}_{h,l}.$$

In the tables below we give the exact and approximate coefficients for two versions of this problem. In \Tref{tab:saws1} are the first 17 exact coefficients $A_L$, followed by 14 approximate coefficients, produced by the method of series extension \cite{G16}.
In \Tref{tab:saws2} we give the corresponding data for the subset of walks that touch all four boundaries, that is $\widehat{A}_{L,L}$.
\begin{table}[htbp!]
   \begin{center}
   \topcaption{SAWs within a square. Approximate coefficients are shown below the line.}
   \begin{tabular}{|l|} \hline
 12\\
 322\\
 14248\\
 1530196 \\
 436619868\\
 343715004510\\
 766012555199052\\
 4914763477312679808\\
 91781780911712980966236\\
 5028368533802124263609489682\\
 813124448051069045700905179168520\\
 389951935424725220050793816994548156972\\
 556688146559188772657829675974686955261012260\\
 2372758431296813754270682928350384434184731056984058\\
 30268116266506218965520430756439292861548616445603740569116\\
 1157901525792517272048285094789582577040591026712499026254697241160\\
 133055180458020027109908469262331157492243786249689003543506387240743574648\\
\hline
4.599082462168733357770670517895290353950$\times 10^{82}$\\
 4.787444735465446541118193916195821352795$\times 10^{91}$\\
 1.502344542130958385779987955512736137038$\times 10^{101}$\\
 1.422486521790516418572011352122294754602$\times 10^{111}$\\
 4.066956104812716452690576247888526246790$\times 10^{121}$\\
 3.513354347231908046171038094563143791710$\times 10^{132}$\\
 9.176123706098699906488924492174438943522$\times 10^{143}$\\
 7.249461481735557442541247664967199805824$\times 10^{155}$\\
 1.733248346210176629051719013852420403142$\times 10^{168}$\\
 1.254591720437060938069650285968046688043$\times 10^{181}$\\
 2.750364377423036423723817745843332088364$\times 10^{194}$\\
 1.826703010696054588130965912464180668109$\times 10^{208}$\\
 3.676760948075648571890631207509538191957$\times 10^{222}$\\
 2.243366976782308538931594331539528235284$\times 10^{237}$\\
\hline
      \end{tabular}
   \label{tab:saws1}
   \end{center}
\end{table}

\begin{table}[htbp!]
   \begin{center}
   \topcaption{SAWs within a square, touching all boundaries of the square. Approximate coefficients are shown below the line.} 
   \begin{tabular}{|l|} \hline
 8\\
 176\\
 9172\\
 1151156\\
 365889264\\
 308813396032\\
 718640738910200\\
 4732971477026278268\\
 89774143793710923662396\\
 4963811733288543976944895936\\
 807029098767544610218003142677260\\
 388251384314275424716358427122233196560\\
 555279544353536878758930497631910438484857672\\
 2369281562338923664687475384593194787991362234069856\\
 30242469017228497029222428170631587362645144066816010779992\\
 1157334830968684997812349313486405793159055139489539075527858732252\\
 133017602531702431926090459378816184891970396066005418878770401897818314324\\
\hline
4.598333987260021227189857734415243170203$\times 10^{82}$\\
 4.786999567406478337697834540691804645783$\times 10^{91}$\\
 1.502268855354498735003841065184804841603$\times 10^{101}$\\
 1.422458373536296174733479079197927114005$\times 10^{111}$\\
 4.066998608403361703661159271340032901512$\times 10^{121}$\\
 3.513533451836503020958527384491093897599$\times 10^{132}$\\
 9.177158624989472547574492769379988707429$\times 10^{143}$\\
 7.250968914970785728403389256700869238767$\times 10^{155}$\\
 1.733854854110844751688740935856711788971$\times 10^{168}$\\
 1.255286461539302161310380869573472578917$\times 10^{181}$\\
 2.752663519800887823692576359393899619149$\times 10^{194}$\\
 1.828921305942165527065491203729824541392$\times 10^{208}$\\
 3.683041623307048542591558080731461688773$\times 10^{222}$\\
 2.248611369730475557411096363398649700261$\times 10^{237}$\\
\hline
      \end{tabular}
   \label{tab:saws2}
   \end{center}
\end{table}

In analysing this data, we make use of the standard methods of series analysis. These include the ratio method and its refinements \cite{GJ09}, the method of differential approximants \cite{GJ09,GJ72} and the method of series extension \cite{G16},
in which the method of differential approximants is used to obtain {\em approximately} a significant number of further terms, based on the known terms. These approximate terms, if of sufficient accuracy, can then be used in the ratio method and its extensions to obtain more precise estimates of the various critical parameters.
A brief summary of all these methods can be found in the appendices of \cite{GJ22}.
In our analysis below we make use of all of these methods.

We are considering SAWs within an $L \times L$ square lattice, with the walks starting anywhere and finishing anywhere within, or on the boundary of, the square.  We also consider a proper subset of those walks that must touch all four edges of the square.

We will be interested in the asymptotic behaviour of $C_L,$ the number of SAWs within an $L \times L$ square, drawn on the square lattice. That is, $C_L = A_L$, as defined above,  when we consider all walks starting anywhere and finishing anywhere within, or on the boundary of, the square, and $C_L=\widehat{A}_{L,L}$ when we consider the proper subset of those walks that must touch all four edges of the square.

The existence of the limit $\lim_{L \to \infty} C_{L}^{1/L^2}=\lambda$ was proved in both \cite{AH78} and \cite{GW90} for SAWs crossing a square. For walks within a square, $A_L$, we proved its existence above. For our proper subset counted by $\widehat{A}_{L,L}$ we have that $R_{L} \leq \widehat{A}_{L,L} \leq A_L$, 
giving the limit's existence in that case. 

This suggests the first method to estimate $\lambda,$ which is to extrapolate the sequence $\lambda_L := C_{L}^{1/L^2}.$ As for the sub-dominant behaviour,
 it is very plausible that the asymptotic behaviour is given by
\begin{equation} \label{eqn:assume}
C_L \sim \lambda^{L^2+bL+c}\cdot L^g,
\end{equation}
as was found for WCAS \cite{GJ22}.
Without, at this point, assuming this supposition, we first extrapolate $\lambda_L$ against $1/L.$ Recall that we only have  17 terms. We used the method of series extension \cite{G16} to extend the sequence of {\em ratios} $r_L \:= C_L/C_{L-1},$ and then use this sequence to extend the
$C_L$ series. In this way we have obtained 14 additional {\em approximate} coefficients. These are given in \Tref{tab:saws1}.

\begin{figure}[ht!] 
\centerline{\includegraphics[width=0.5\textwidth,angle=0]{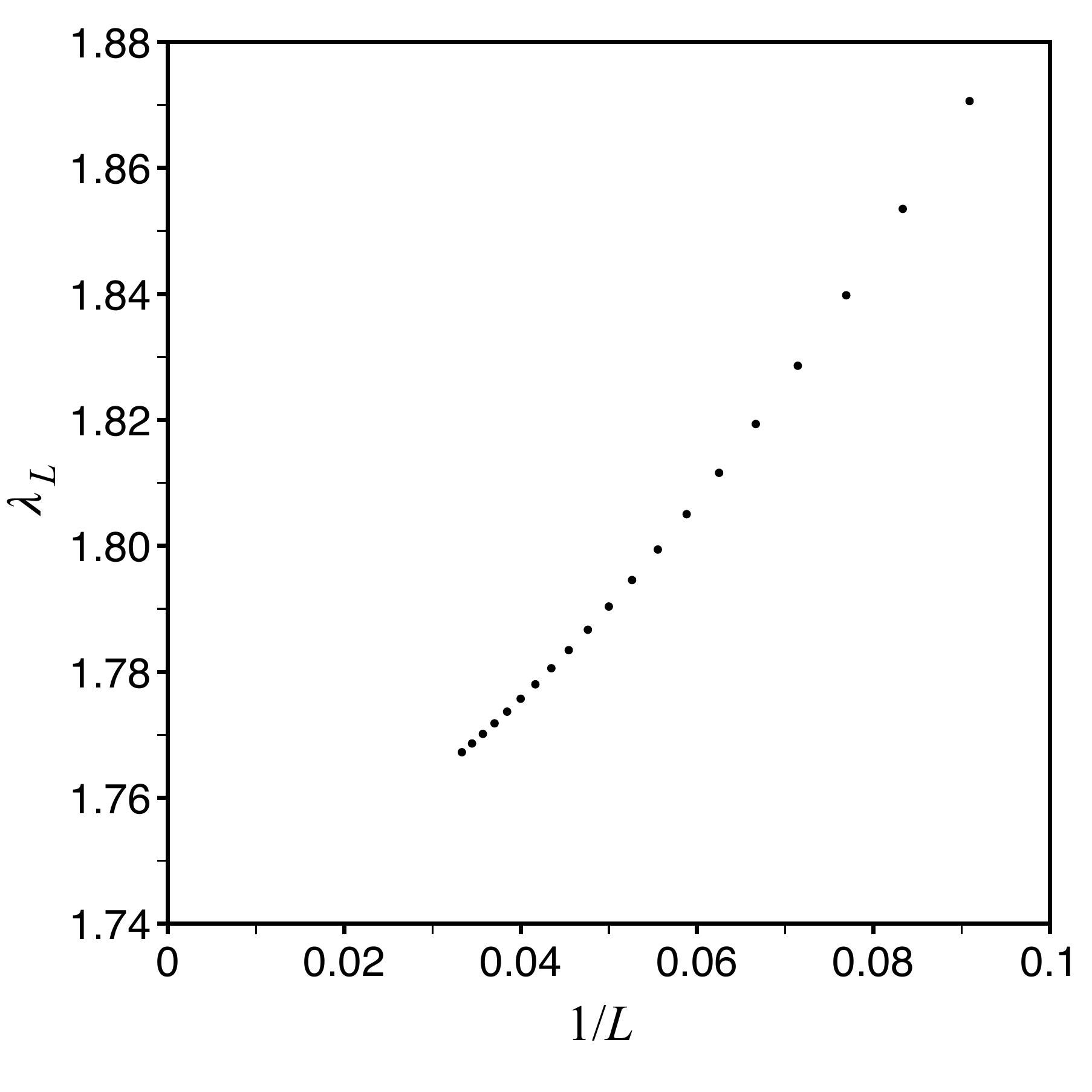} }
\caption{$\lambda_L$ plotted against $1/L$.} 
\label{fig:lam1}
\end{figure}

We show a plot of $\lambda_L$ against $1/L$ in \Fref{fig:lam1}. It displays considerable curvature, though can be visually extrapolated to $\lambda \approx 1.75.$ 

\begin{figure}[ht!] 
\begin{minipage}[t]{0.45\textwidth} 
\centerline{\includegraphics[width=\textwidth,angle=0]{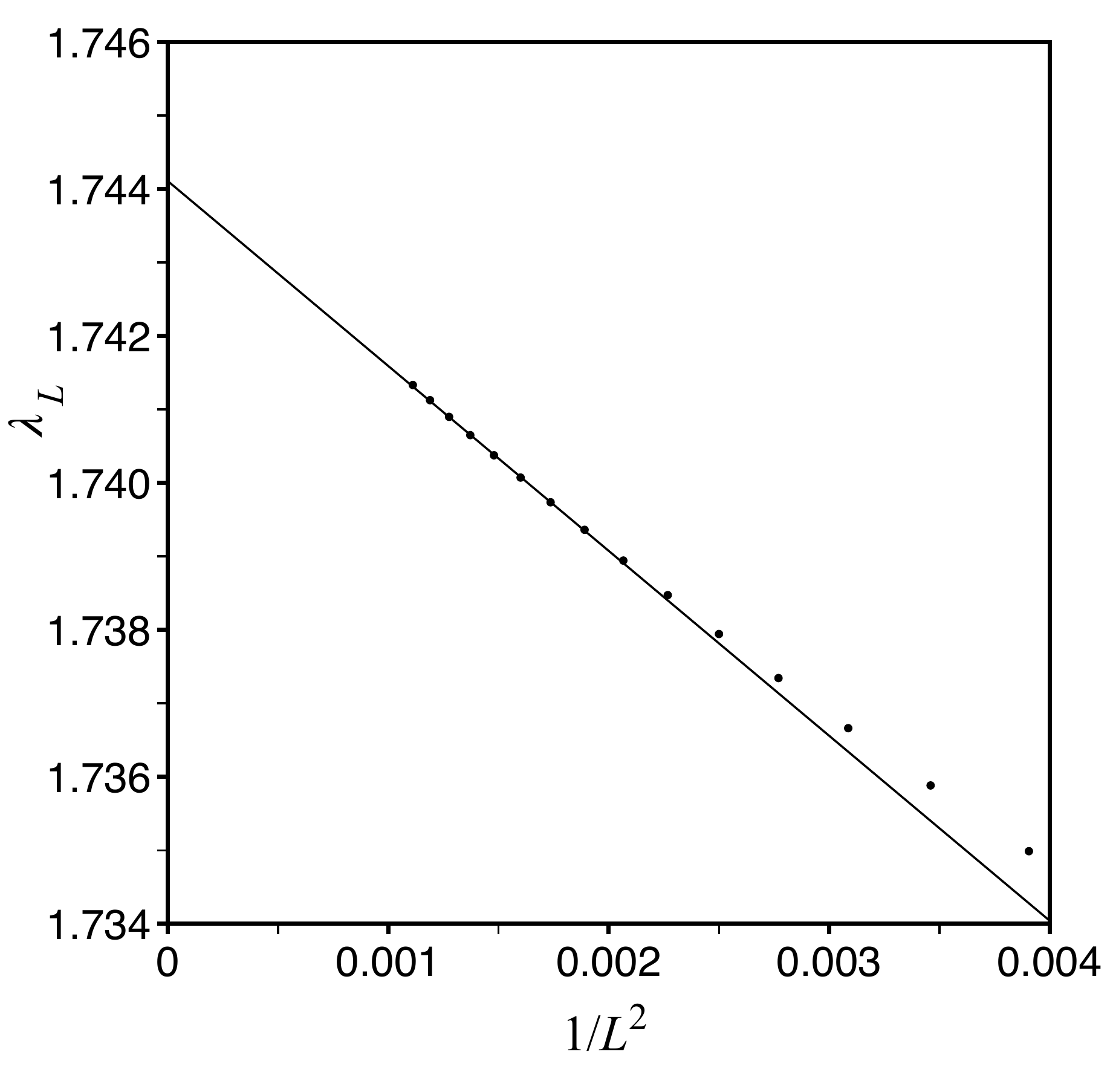} }
\end{minipage}
\hspace{0.05\textwidth}
\begin{minipage}[t]{0.45\textwidth} 
\centerline{\includegraphics[width=\textwidth,angle=0]{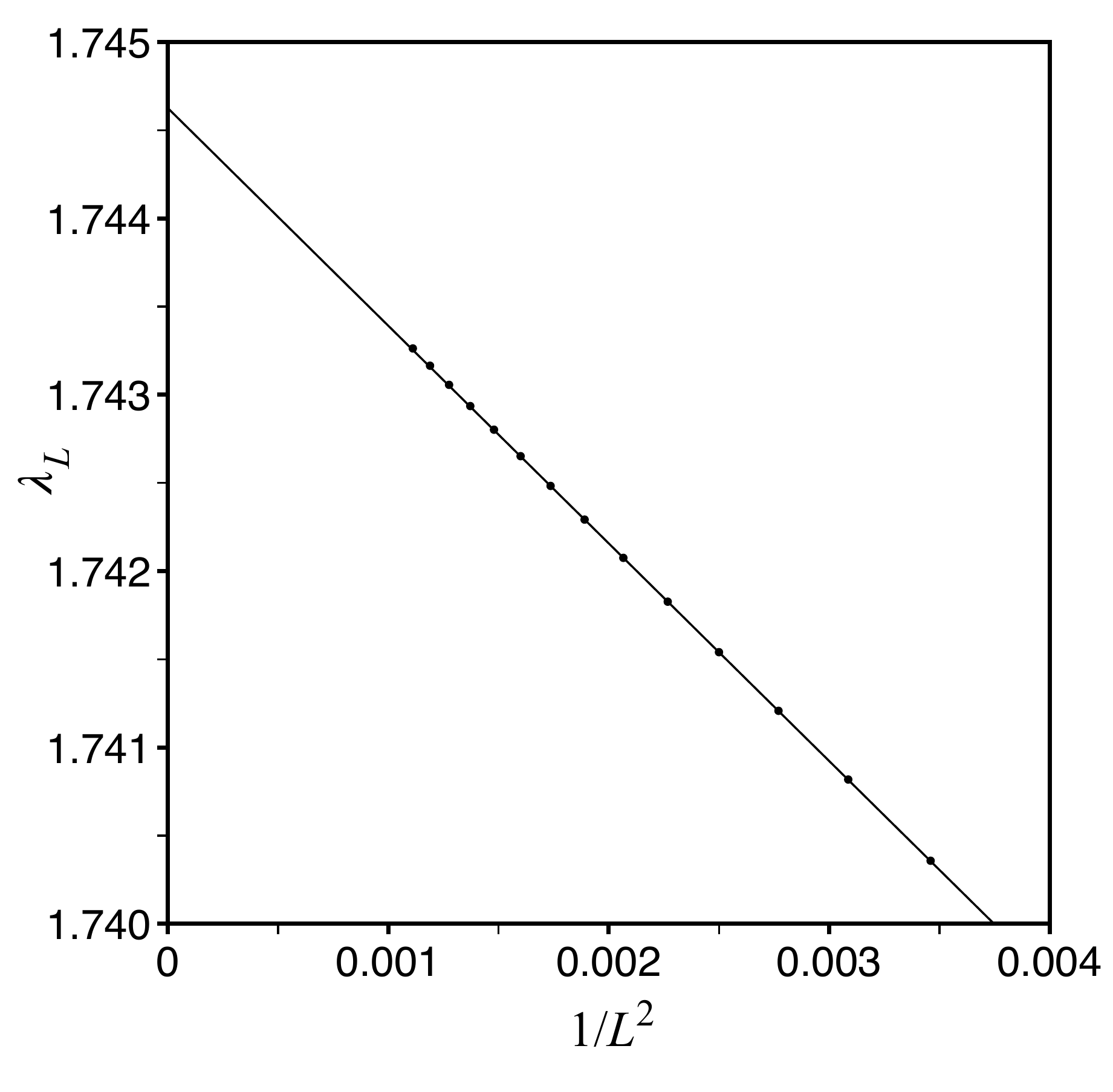}}
\end{minipage}
\caption{\label{fig:lam2} Values of estimators $\lambda_L$ plotted against $1/L^2$ from quadratic 
fits (left panel) and cubic fits (right panel). The intercepts happen at 1.74411 and 1.74462, respectively.}
\end{figure}

It is reasonable to assume that the curvature is due to the presence of higher-order terms, such as $1/L^2,$ $1/L^3$ etc. In  the left panel of \Fref{fig:lam2} we show the values of the estimators $\lambda_L$ of $\lambda$ assuming $\lambda_L$ converges to $\lambda$ with correction terms $c_1/L + c_2/L^2,$ plotted against $1/L^2.$ Similarly in the right panel of \Fref{fig:lam2} we show the values of the estimators of $\lambda$ assuming $\lambda_L$ converges with correction terms $c_1/L + c_2/L^2+c_3/L^3,$ plotted against $1/L^2.$ In each plot we also include as a guide to the eye a linear
fit to the data.

\begin{figure}[ht!] 
\begin{minipage}[t]{0.45\textwidth} 
\centerline{\includegraphics[width=\textwidth,angle=0]{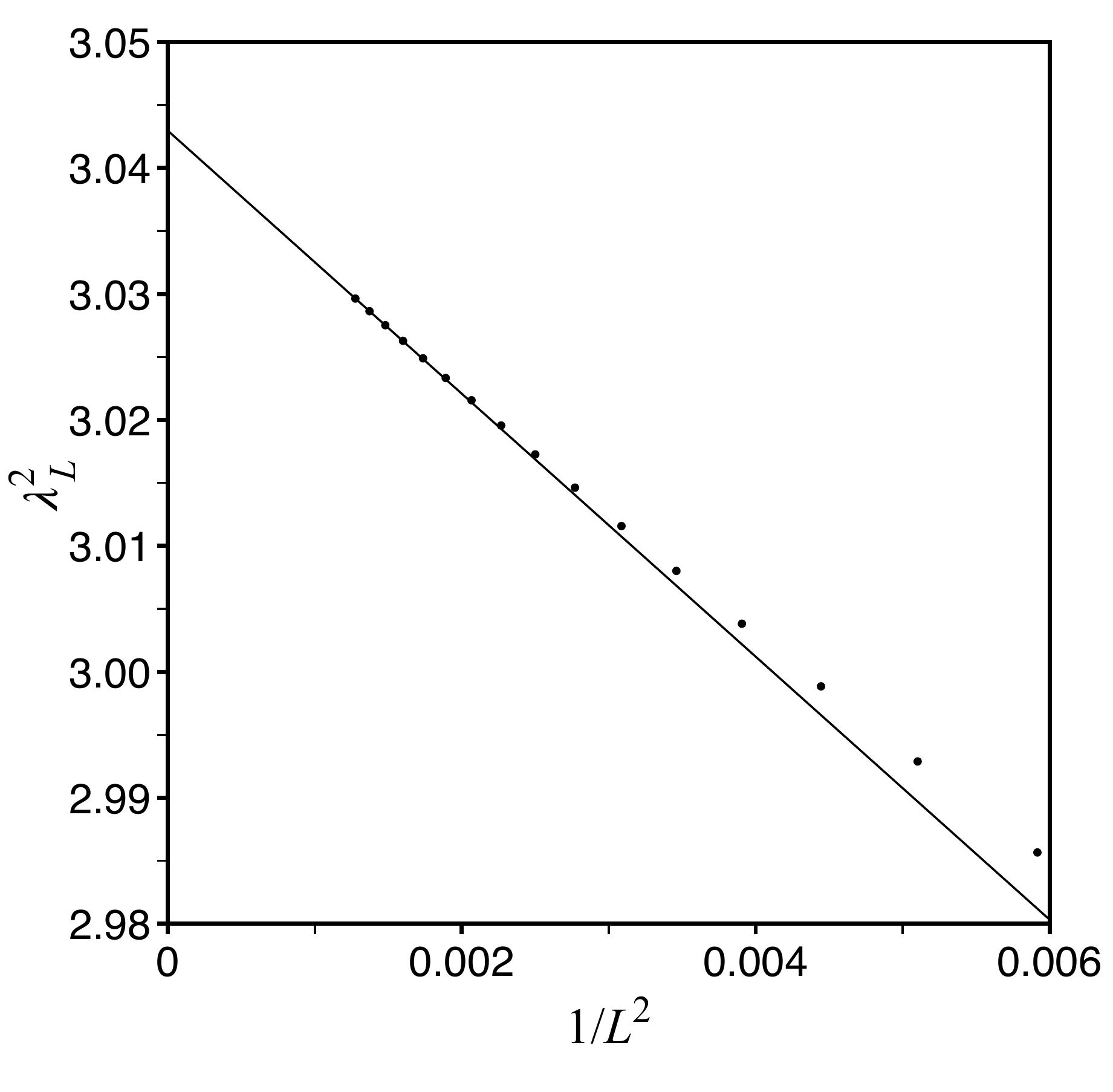} }
\end{minipage}
\hspace{0.05\textwidth}
\begin{minipage}[t]{0.45\textwidth} 
\centerline{\includegraphics[width=\textwidth,angle=0]{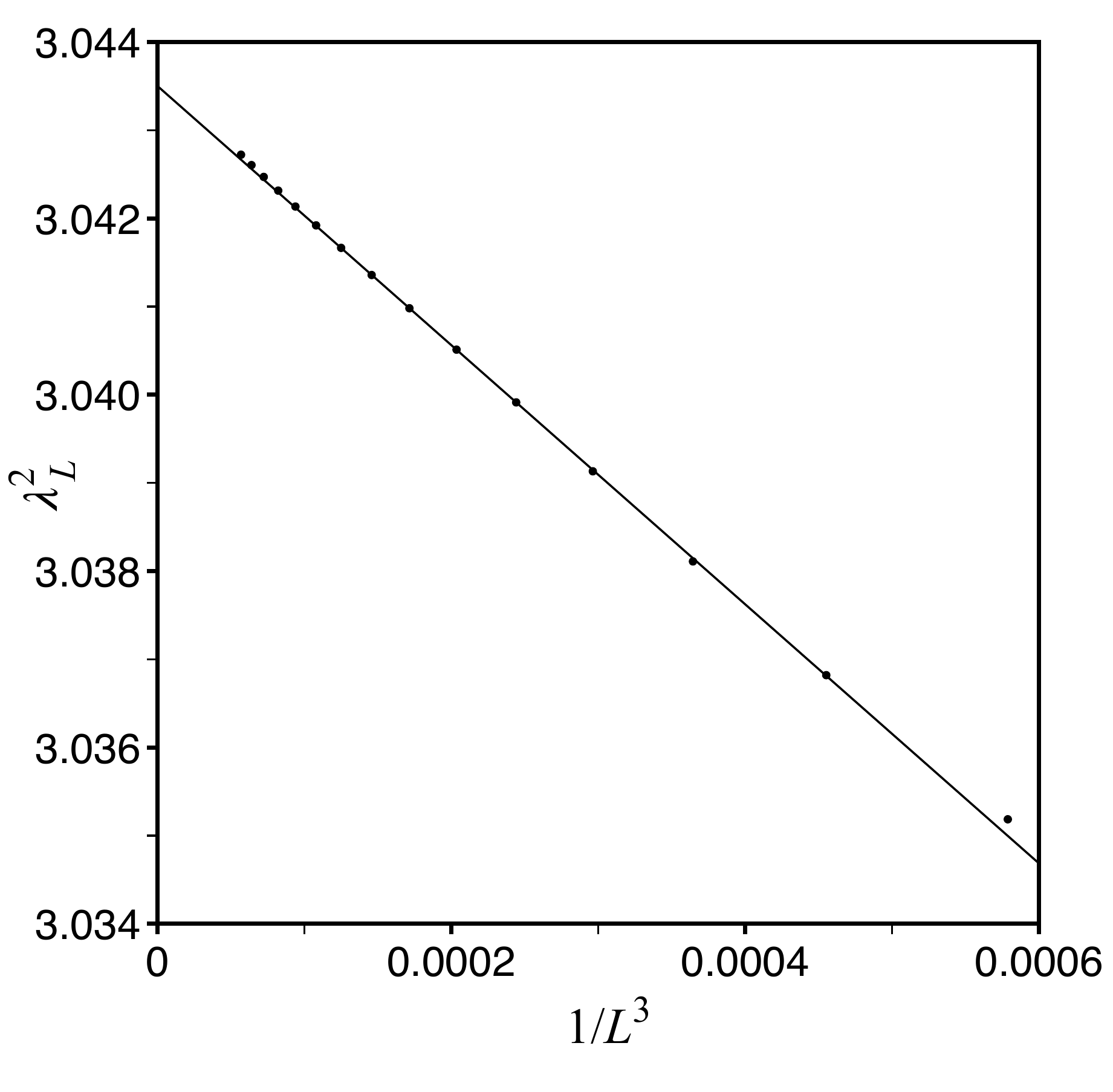}}
\end{minipage}
\caption{\label{fig:lsq} Estimates of $\lambda_L^2$ from ${\mathcal C}_L$ plotted against $1/L^2$ (left panel) and from a quadratic fit plotted against $1/L^3$ (right panel). The intercepts happen at
3.04395 and 3.04350, respectively}
\end{figure}

While it has not been proved that the ratios $R_L \equiv C_L/C_{L-1} \sim \lambda^{2L},$ it is almost certainly true, and we will assume it to be so in our subsequent analysis. There are several numerical methods available to estimate the value of $\lambda.$ The first of these is to extrapolate the ratios of these ratios against $1/L$ and then against higher order polynomials in $1/L,$ as we did for the sequence $C_{L}^{1/L^2}$ above. 
Given the expectation that $R_L \sim \lambda^{2L},$ we define the ratio of ratios 
$${\mathcal C}_L :=  \frac{R_{L+1}}{R_L}= \frac{C_{L+1}\cdot C_{L-1}}{C_L^2}.$$ From (\ref{eqn:assume}) it follows that 
\begin{equation}
{\mathcal C}_L = \lambda^2 \left ( 1 -\frac{g}{L^2} + O(L^{-3}) \right ).
\end{equation}
Accordingly, we fit the sequence $\{{\mathcal C}_L\} $ to $c_1 + c_2/L^2,$ so that $c_1$ should give estimators of $\lambda^2,$ and $c_2$ should give estimators of $-\lambda^2\cdot g.$ In \Fref{fig:lsq} we show plots of ${\mathcal C}_L$ vs $1/L^2$ and  the values $\lambda_L^2$ of the estimators of $\lambda^2$ from a quadratic fit vs $1/L^3$. From this we estimate $\lambda^2 = 3.0435 \pm 0.0002,$ or $\lambda = 1.74456 \pm 0.00006$. 

Both methods produce  estimates entirely consistent with the value of $\lambda_S,$ the corresponding growth constant for walks crossing a square.

Returning now to the question of the sub-dominant terms as per equation (\ref{eqn:assume}), 
we will attempt to estimate the parameters $b,$ $c$ and $g$ in this assumed form by three different methods. 

\begin{figure}[ht!] 
 \centerline{\includegraphics[width=0.5\textwidth,angle=0]{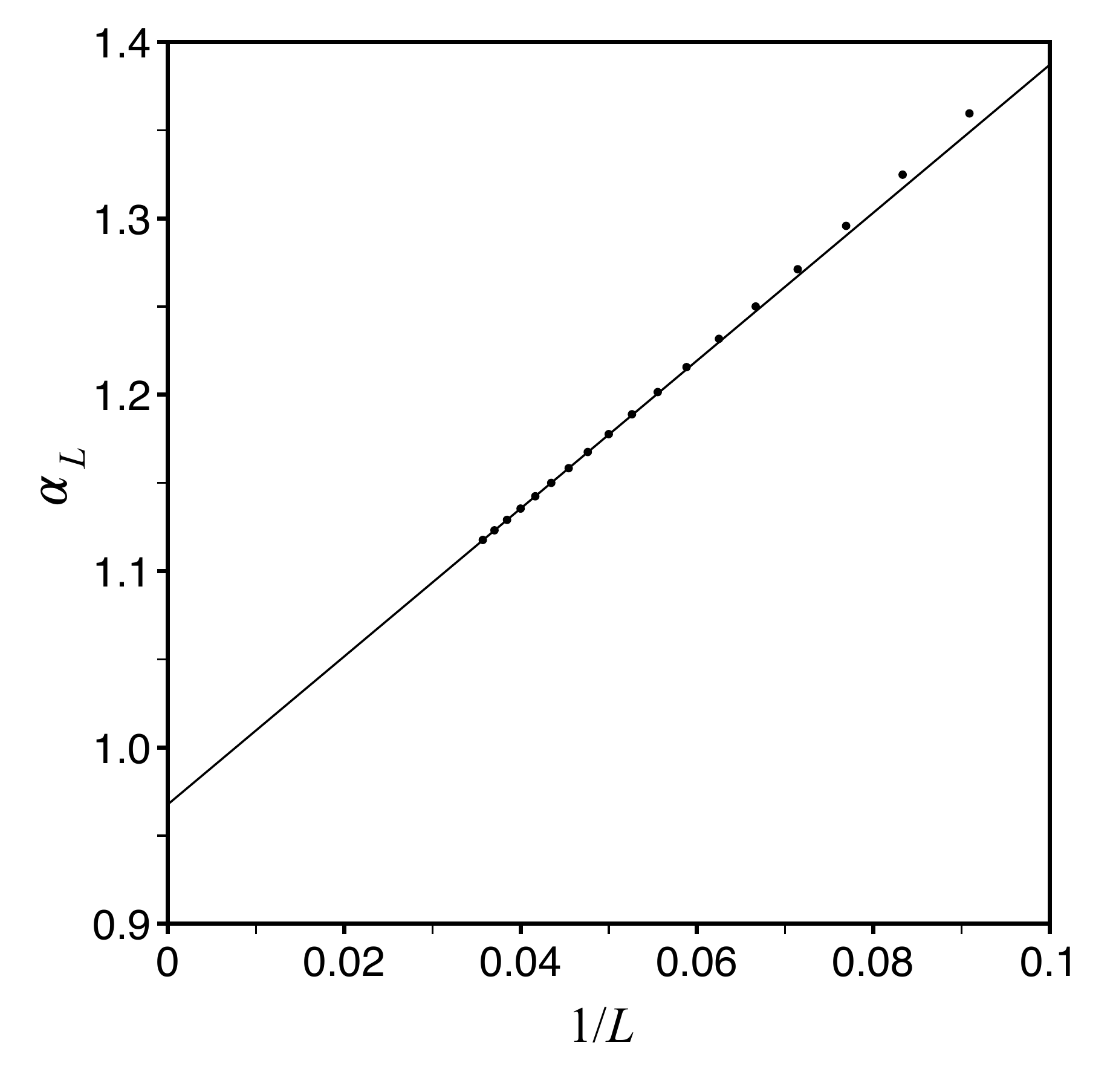} }
\caption{Ratios $\alpha_L=d_L/d_{L-1} \sim \alpha$ plotted against $1/L.$} 
\label{fig:alpha1}
 \end{figure}

Using our best estimate of $\lambda=\lambda_S=1.7445498 \pm 0.0000012$ from \cite{GJ22} we first form the sequence $d_L := C_L/\lambda^{L^2} \sim \lambda^{bL+c}L^g.$ If the assumed form is correct this sequence behaves as a typical power-law singularity, in which the coefficients grow as $a_n \sim C \alpha^n \cdot n^g,$ and can be analysed as such. With that notation, the growth constant $\alpha = \lambda^b,$ and the amplitude $C = \lambda^c.$ Of course, we have to use the estimated value $\lambda=1.7445498$.
In \Fref{fig:alpha1} we show the ratios $\alpha_L:= d_L/d_{L-1} \sim \alpha$ plotted against $1/L.$  We estimate from this plot that $\alpha = 0.975 \pm 0.025,$ so that $b=\log \alpha/\log \lambda \approx -0.045 \pm 0.05,$ suggesting that the term $\lambda^{bL}$ is small or absent. The gradient gives an estimate of the exponent $g$, and in this way we estimate $g=4.0 \pm 0.5.$ Our estimates of $b$ and $g$ are too imprecise to estimate the value of $c.$ 
The estimates of $\alpha$ and $b$ are conservative enough to encompass the full range of uncertainty in the estimate of $\lambda$.

\begin{figure}[ht!] 
\begin{minipage}[t]{0.45\textwidth} 
\centerline{\includegraphics[width=\textwidth,angle=0]{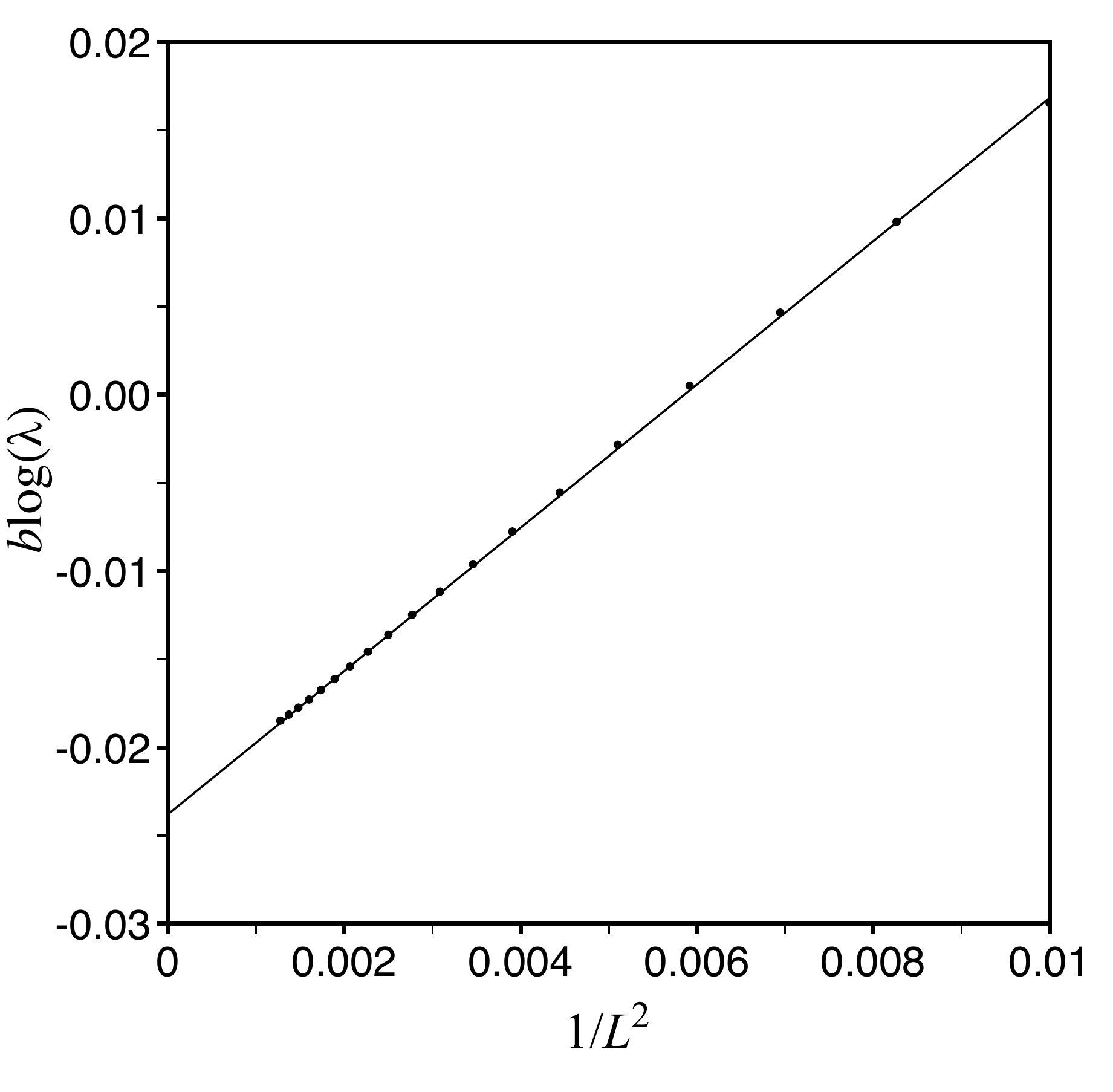}}
\end{minipage}
\hspace{0.03\textwidth}
\begin{minipage}[t]{0.45\textwidth} 
\centerline{\includegraphics[width=\textwidth,angle=0]{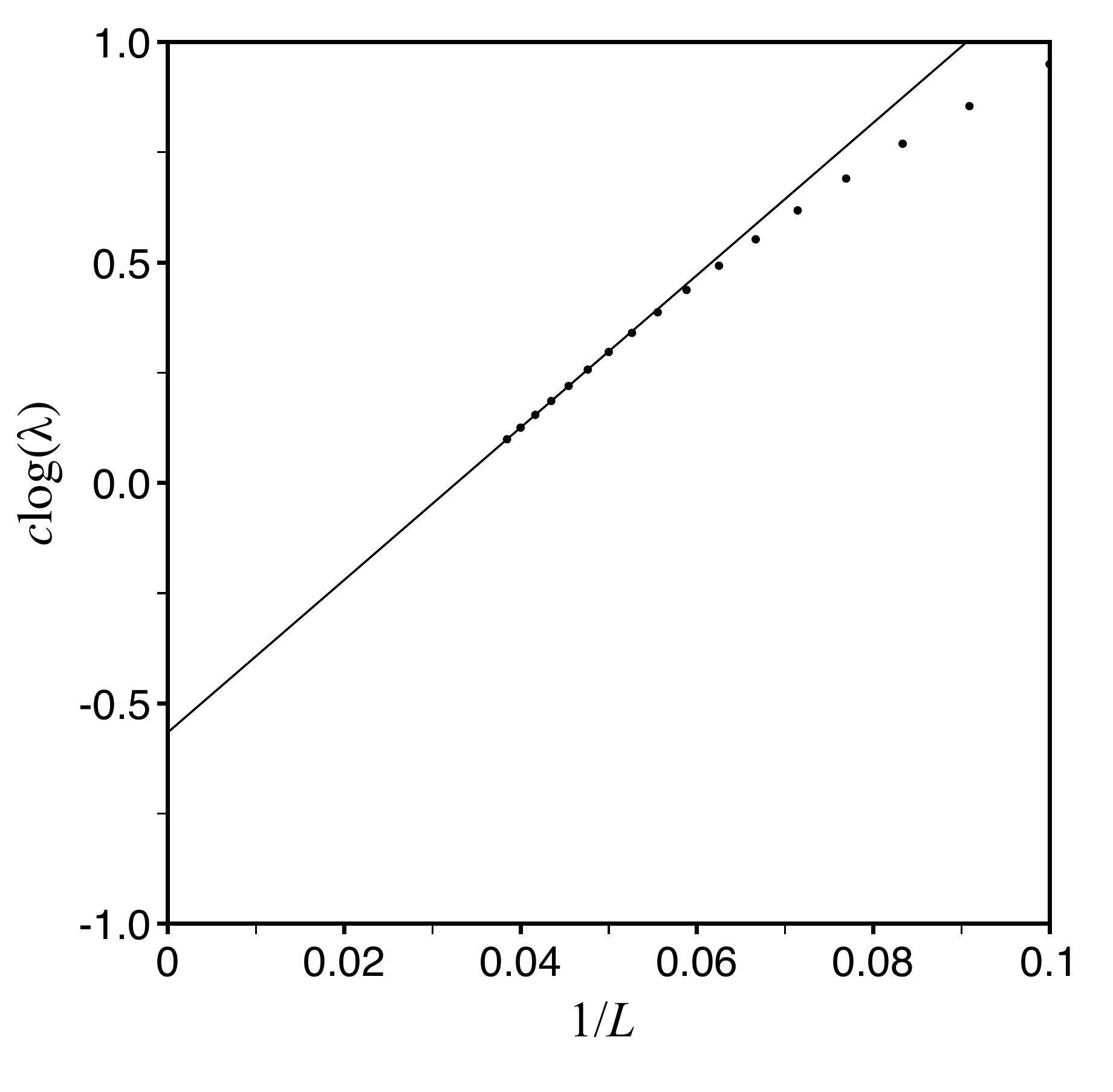} }
\end{minipage}

 \begin{minipage}[t]{0.45\textwidth} 
\centerline{\includegraphics[width=\textwidth,angle=0]{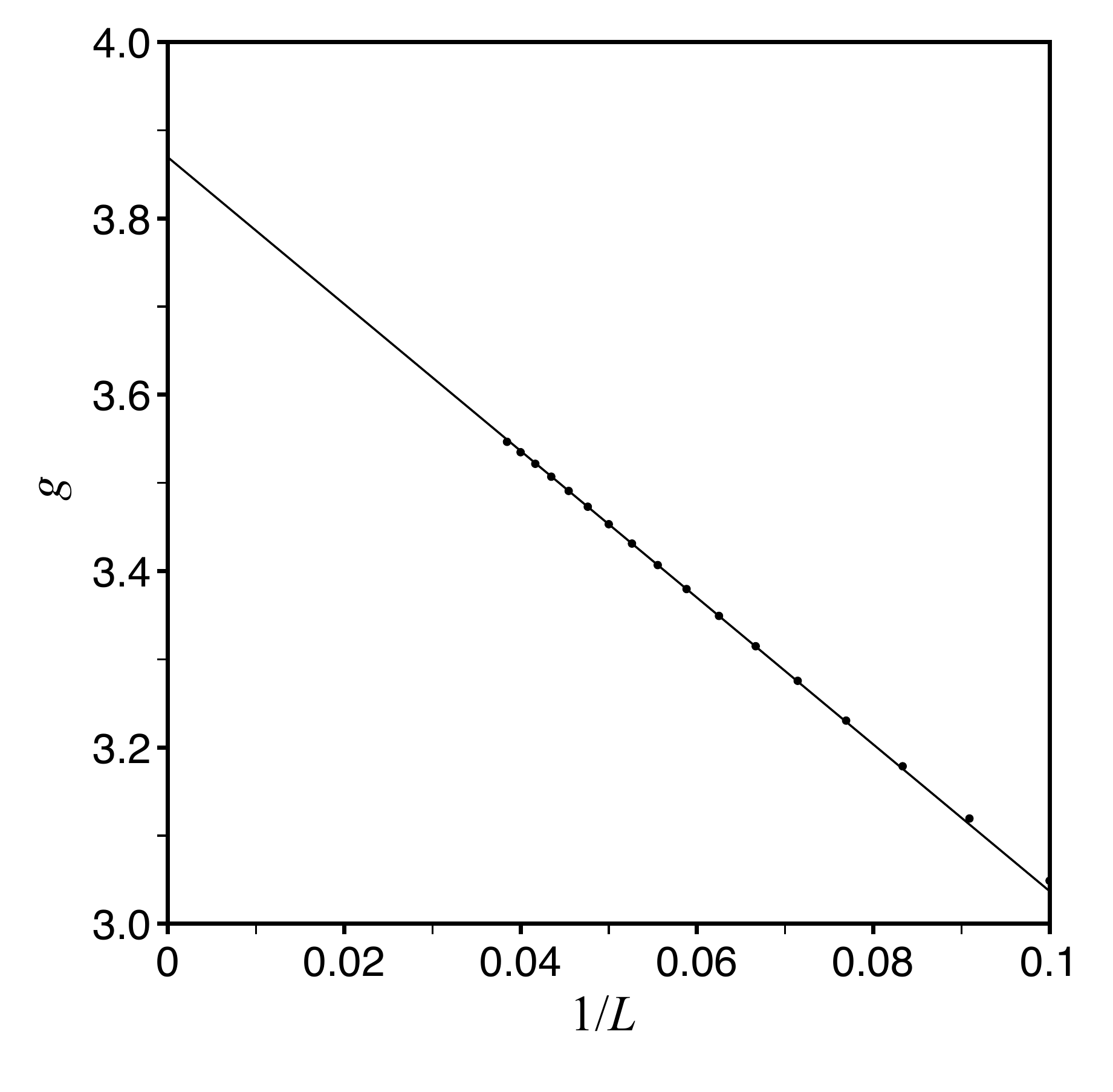}}
\end{minipage}
\hspace{0.03\textwidth}
 \begin{minipage}[t]{0.45\textwidth} 
\centerline{\includegraphics[width=\textwidth,angle=0]{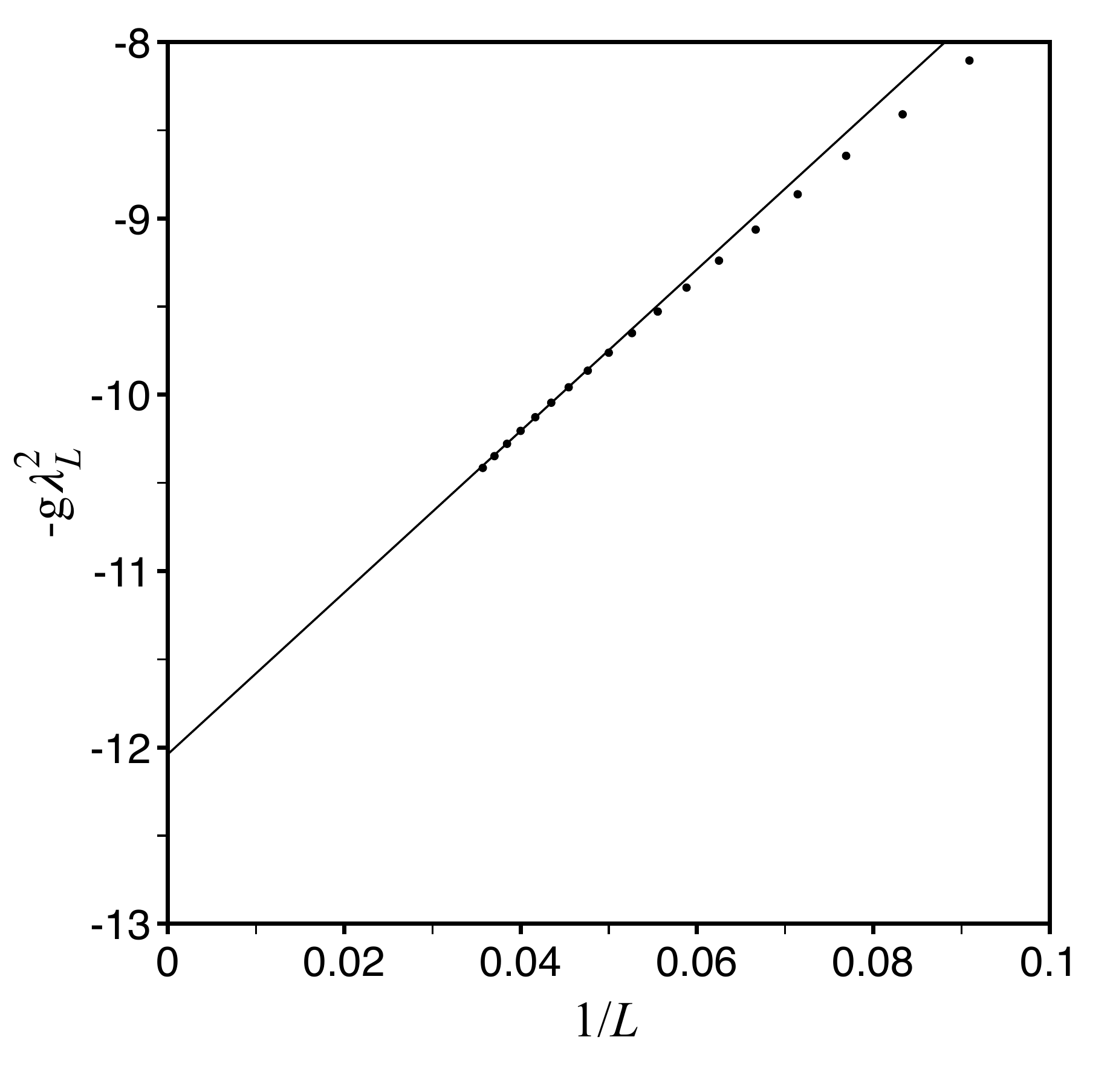}}
\end{minipage}
\caption{\label{fig:para} Values of the estimators of $b \log{\lambda},$ $c \log{\lambda},$ $g$, and $-g\lambda_L^2$.}
\end{figure}

The second method used is to fit to the assumed form by writing $$\log{d_L} \sim b\log(\lambda)  L + c \log(\lambda) + g \log{L}.$$ We then use successive triples of values $( \log{d_{k-1}},\log{d_{k}},\log{d_{k+1}}),$ with $k=2,3,\cdots, L_{\textrm{max}}-1,$ to give estimates of the parameters $b \log{\lambda},$ $c \log{\lambda},$ and $g.$ Plots of these against various powers of $1/L$ are shown in \Fref{fig:para} as is a plot $-g\lambda_L^2$ obtained from the quadratic fit to
the sequence $\{{\mathcal C}_L\}$.

From these plots, we estimate $b\log{\lambda} = -0.024 \pm 0.003,$ or $b = -0.043 \pm 0.005,$ $c \log{\lambda} = -0.75 \pm 0.25,$ or $c = -1.35 \pm 0.45,$ and $g \approx 3.8$. The agreement between the two methods for the estimate of $g$ is quite good, and well within quoted confidence limits. 

We can however do better by comparing this data with the WCAS data discussed in \cite{GJ22}. We have substantially longer series there, and we know that the growth constant $\lambda$ is the same as for the walks we are considering here. For the other parameters the estimates are $b = -0.04354 \pm 0.00001,$ $c = 0.5624 \pm 0.0005,$ and $g = 0.000 \pm 0.005.$

Note that the value of $b$ is similar, and it seems at least plausible that it is the same. To investigate this further, we form the Hadamard quotient of the two sequences, the WCAS sequence and this one. This involves taking the coefficient by coefficient quotient, which eliminates the leading-order term, $\lambda^{L^2}.$ Indeed, we expect the quotient coefficients to behave like a power-law series, so with coefficients growing like $C \cdot \mu^L \cdot L^h,$ where if the value of $b$ for the two series is the same, then $\mu=1,$ and the exponent $h$ should be the difference between the two exponents, $g$ for WCAS and $g$ for walks within a square.

We analysed this quotient sequence by the method of differential approximants, and obtained $\mu=1.0000 \pm 0.0002,$ and $g = 3.9 \pm 0.1$. This provides strong evidence for the conjecture that the two sub-dominant terms are the identical, and that the exponent for walks within a square is $g=3.9 \pm 0.1$ The estimate of the amplitude $C$ is very sensitive to the value of $g,$ varying between 3 and 5 as $g$ varies within the quoted range. The estimates of $c$ given above for the two problems are consistent with this range, but we cannot make any more precise estimate of $c$ using this method.

Accordingly we conclude this section with the conjecture that $$ C_L \sim \lambda^{L^2+bL+c}\cdot L^g,$$ where $\lambda = \lambda_S$ (our best estimate \cite{GJ22} is  $\lambda_S=1.7445498$), $b=-0.04354 \pm 0.00001,$ $c=-1.35 \pm 0.45,$ and $g=3.9 \pm 0.1.$

We repeated this analysis for the proper subset of walks that touch all four sides of the square. The relevant series are listed in \Tref{tab:saws2}. The results were very similar to those for we have just analysed, though slightly less well behaved. Our central estimates of all parameters would be similar, but with greater uncertainties. A comparison of the coefficients in Tables \ref{tab:saws1} and \ref{tab:saws2} makes it clear why this is the case. By order 15 the coefficients agree to the first 4 or 5 significant digits, and this persists. Accordingly, we conjecture that these behave identically, asymptotically.

To investigate this further, we studied the Hadamard quotient of the two series, and found that they were indeed approaching the expected limit of 1, indeed, using the last known coefficients, the quotient is $0.99971757\cdots.$ A ratio plot shows that this ratio sequence approaches 1 with zero gradient, which implies that the sub-dominant terms $b,$ $c$ and $g$ are the same.

Finally, we present our analysis of the enumeration data $P_L$ for self-avoiding polygons contained in square of size $L$ (also commonly known as the problem of enumerating cycles in the square grid). Iwashita et. al computed the number of cycles on square grids up to size $L=26$ (see \cite{INK13} for a description of their algorithm). The data can be found on the OEIS as sequence \seqnum{A140517}. Since this is a substantially longer series than the previous two, we were able to extend the $P_L$ series by 30 additional approximate coefficients. 

\begin{figure}[ht!] 
\begin{minipage}[t]{0.45\textwidth} 
\centerline{\includegraphics[width=\textwidth,angle=0]{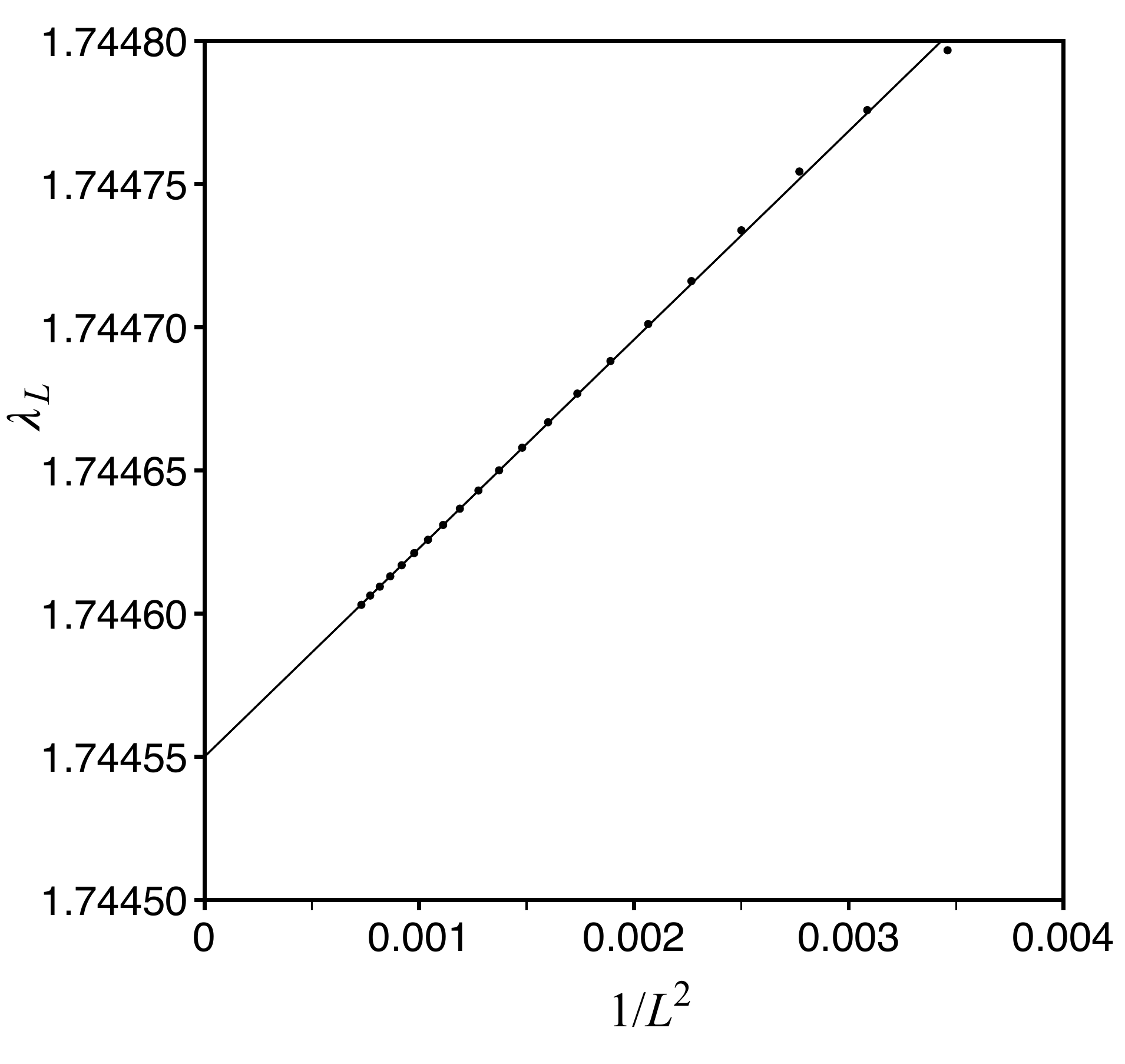}}
\end{minipage}
\hspace{0.03\textwidth}
\begin{minipage}[t]{0.45\textwidth} 
\centerline{\includegraphics[width=\textwidth,angle=0]{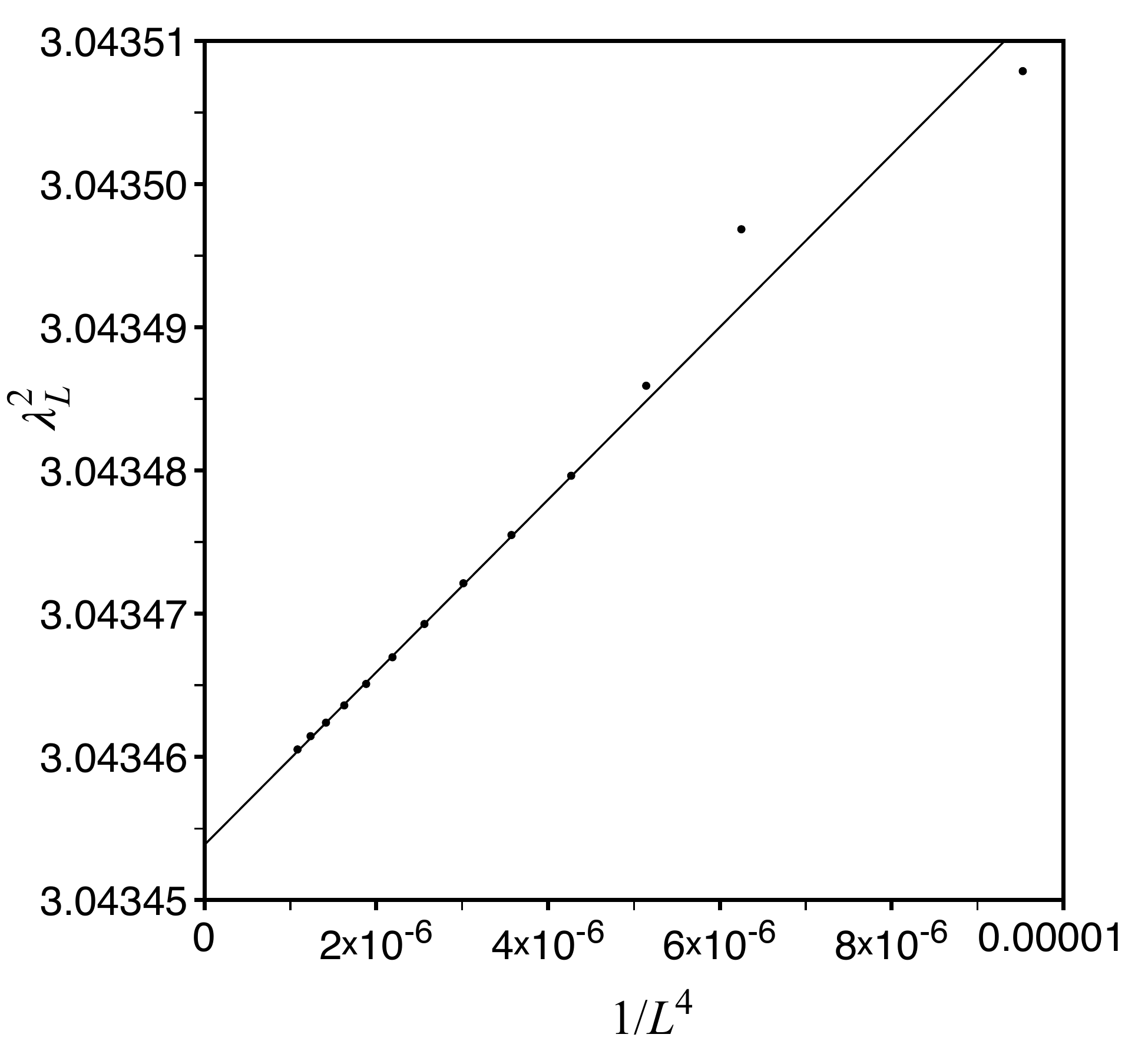} }
\end{minipage}
\caption{Values of the estimators $\lambda_L$ and $\lambda_L^2$ for cycles in the square grid.} 
\label{fig:sap-lambda}
 \end{figure}

Firstly, we estimated $\lambda$ by the two methods described above. In the left panel of \Fref{fig:sap-lambda} we show a plot of $\lambda_L:= P_L^{1/L^2}$ after fitting to a quartic  correction term in $1/L$ ($c_1/L+c_2/L^2+c_3/L^3+c_4/L^4$), while in the right panel we show a plot
of the values of the estimator $\lambda_L^2$ obtained after fitting ${\mathcal P}_L:=(P_{L+1}P_{L-1})/P_L^2$ to a cubic of the form $c_0+c_2/L^2+c_3/L^3$. 
The first plot gives us the estimate $\lambda \approx 1.744550$, while the second plot 
has an intercept at 3.043456 corresponding to $\lambda \approx 1.7445504$. Both of these confirm to a high degree of accuracy that $\lambda=\lambda_S$.

\begin{figure}[ht!] 
\begin{minipage}[t]{0.45\textwidth} 
\centerline{\includegraphics[width=\textwidth,angle=0]{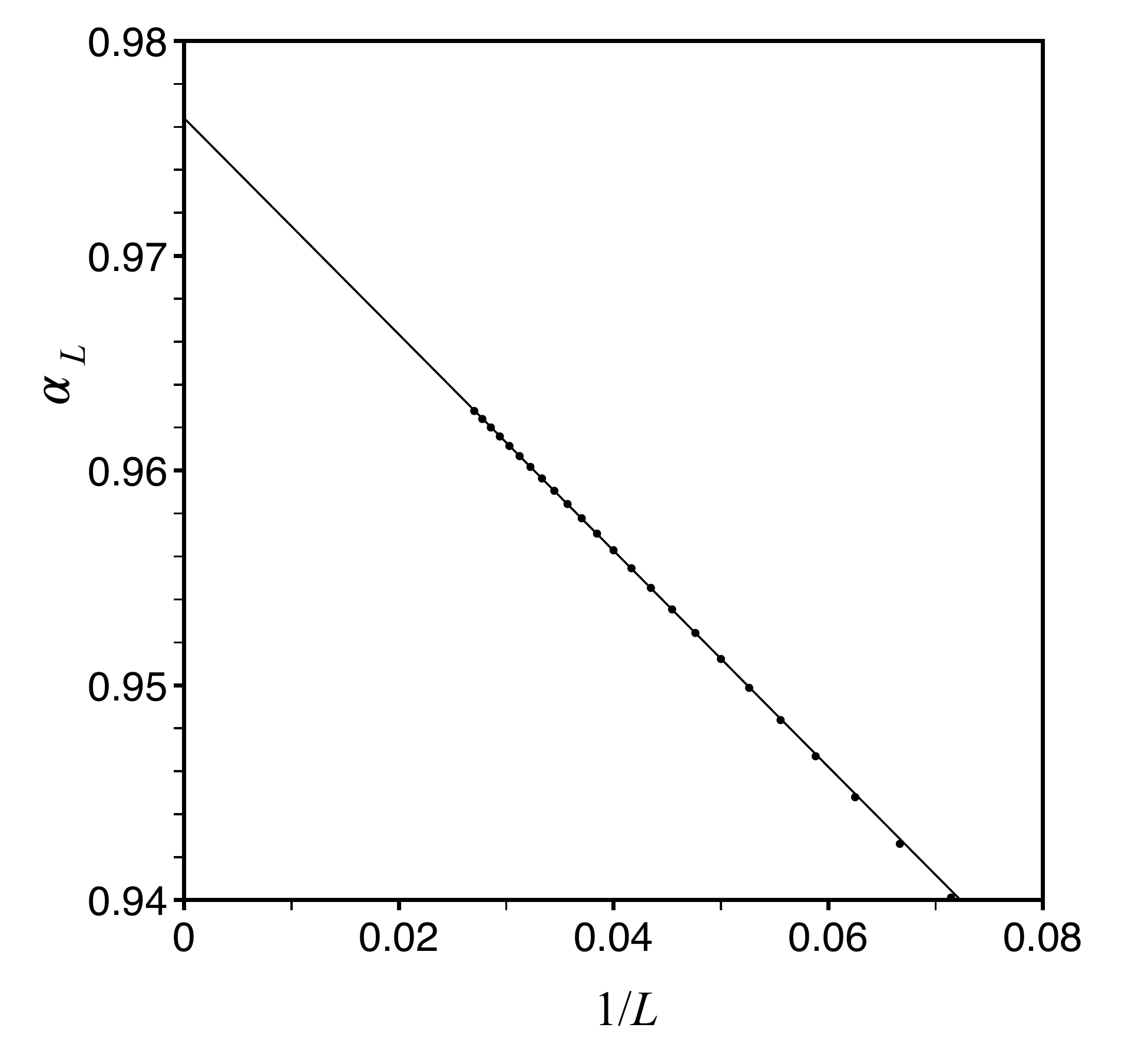}}
\end{minipage}
\hspace{0.03\textwidth}
\begin{minipage}[t]{0.45\textwidth} 
\centerline{\includegraphics[width=\textwidth,angle=0]{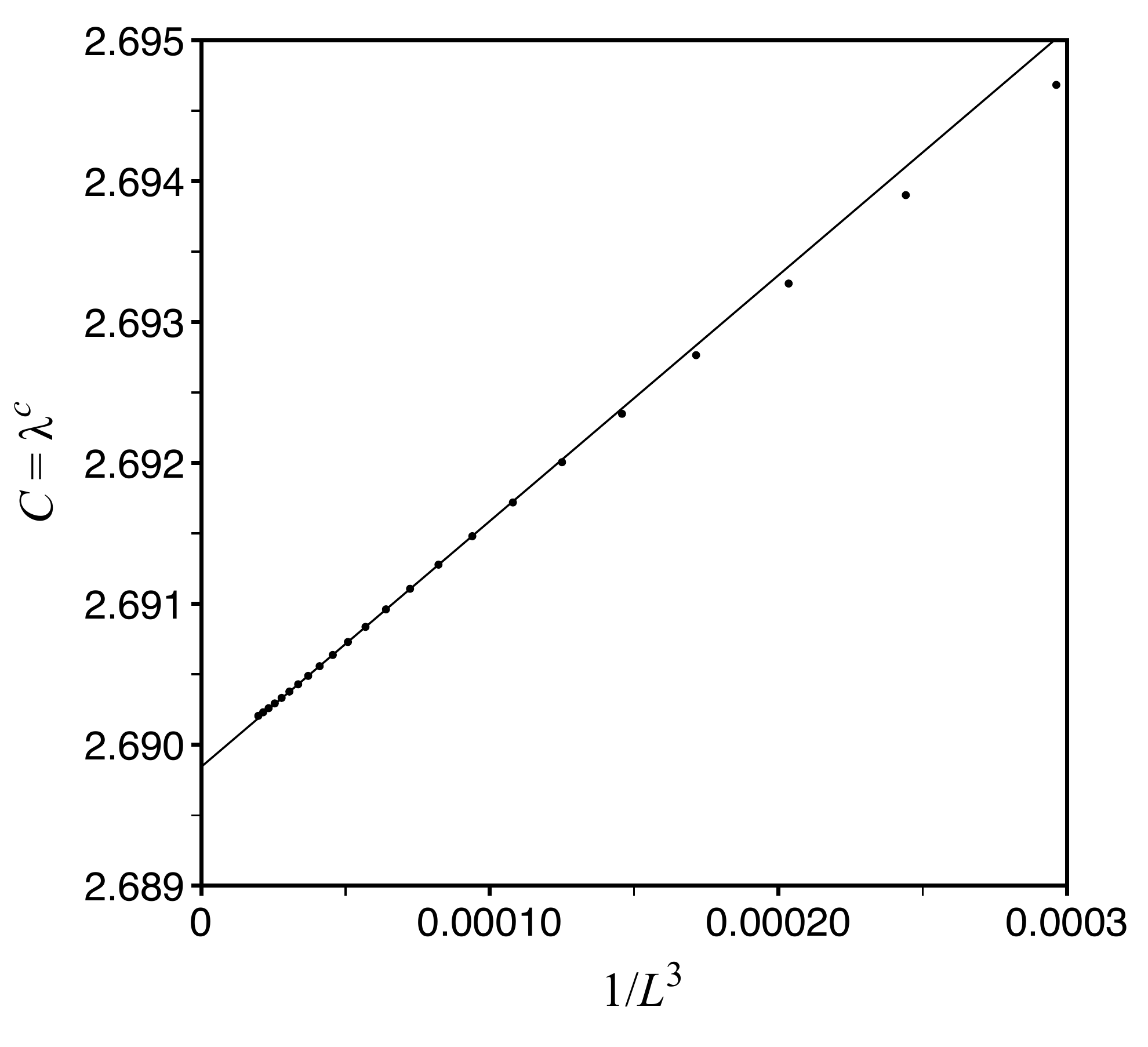} }
\end{minipage}
\caption{Ratios $\alpha_L=d_L/d_{L-1} \sim \alpha$ plotted against $1/L$ and 
$C=d_L/(\lambda^{bL}L^g) \sim \lambda^c$ plotted against $1/L^3$ for cycles in a square grid. } 
\label{fig:sap-alpha}
 \end{figure}

Secondly, we estimate the parameter $b,c$, and $g$ in the assumed asymptotic form 
$$P_L \sim \lambda^{L^2+bL+c}\cdot L^g.$$ 
In the left panel of  \Fref{fig:sap-alpha} we show the ratios $\alpha_L:= d_L/d_{L-1} \sim \alpha$ plotted against $1/L.$  We estimate from this plot that $\alpha = 0.976 \pm 0.002$. It is therefore  reasonable to assume that the value of $\alpha$ for cycles equals that for WCAS where we have the more accurate estimate $\alpha = 0.97606\pm 0.00001$. We shall use this value in our subsequent analysis.
The gradient gives an estimate of the exponent $g$, and in this way we estimate $g=-0.500 \pm 0.005.$ We thus conjecture that $g=-\frac12$, exactly. In the right panel of \Fref{fig:sap-alpha} we show the values of $C=d_L/(\lambda^{bL}L^g) \sim \lambda^c$ plotted against $1/L^3$. From this plot we estimate that $\lambda^c = 2.690 \pm 0.001$ and hence $c=1.7782\pm0.0006$. Note that this method for estimating $c$ is extremely sensitive to the values one uses for $\lambda$ and $b$ (assuming $g$ is known exactly), so while the estimate for $c$ is quite accurate we are not confident that the error-bars can be trusted. The interested reader can make use of the data and Maple worksheet at our GitHub repository (see Section~\ref{sec:resource}) to further explore these numerical issues.

\begin{figure}[ht!] 
\begin{minipage}[t]{0.45\textwidth} 
\centerline{\includegraphics[width=\textwidth,angle=0]{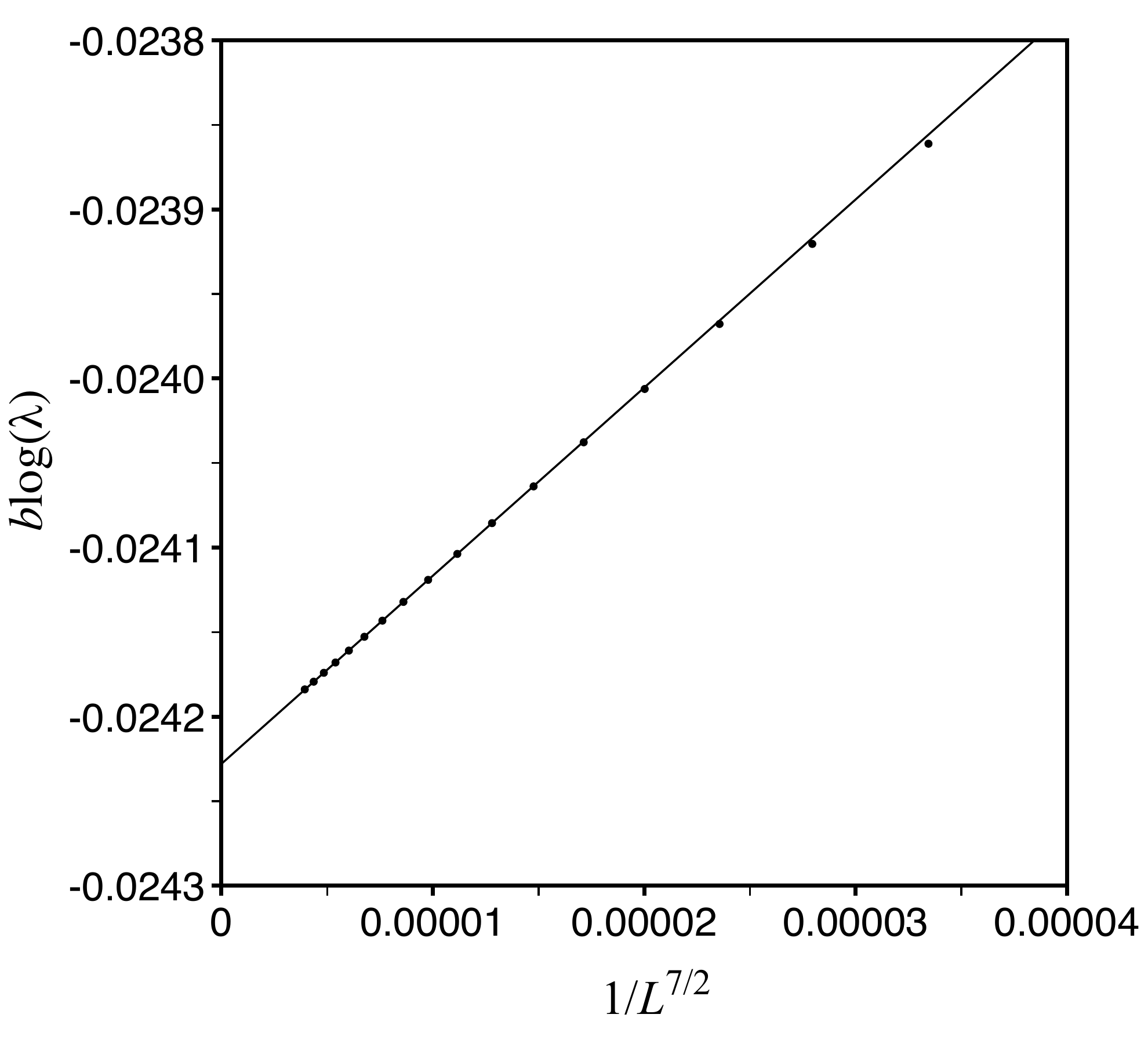}}
\end{minipage}
\hspace{0.03\textwidth}
\begin{minipage}[t]{0.45\textwidth} 
\centerline{\includegraphics[width=\textwidth,angle=0]{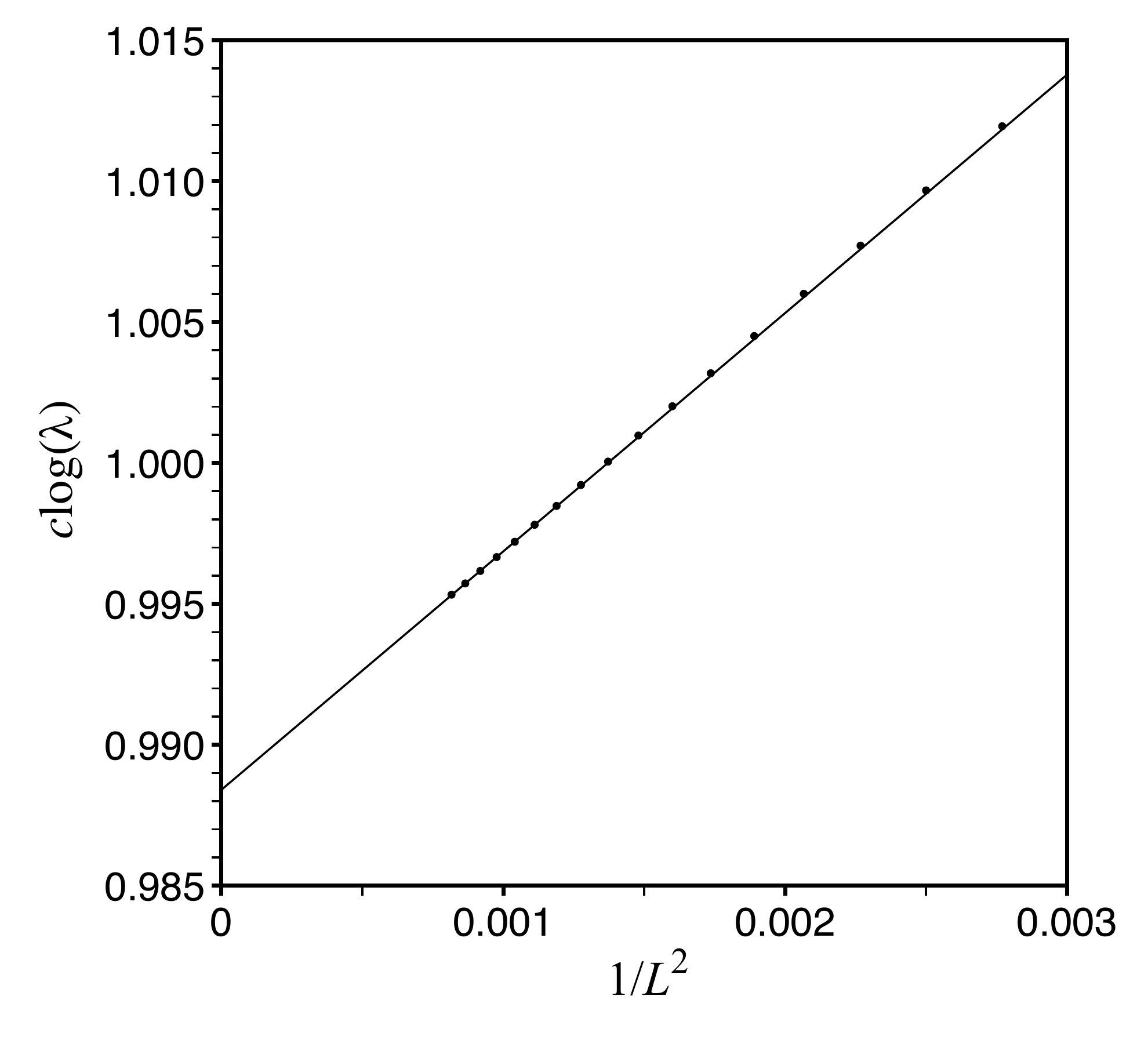} }
\end{minipage}

 \begin{minipage}[t]{0.45\textwidth} 
\centerline{\includegraphics[width=\textwidth,angle=0]{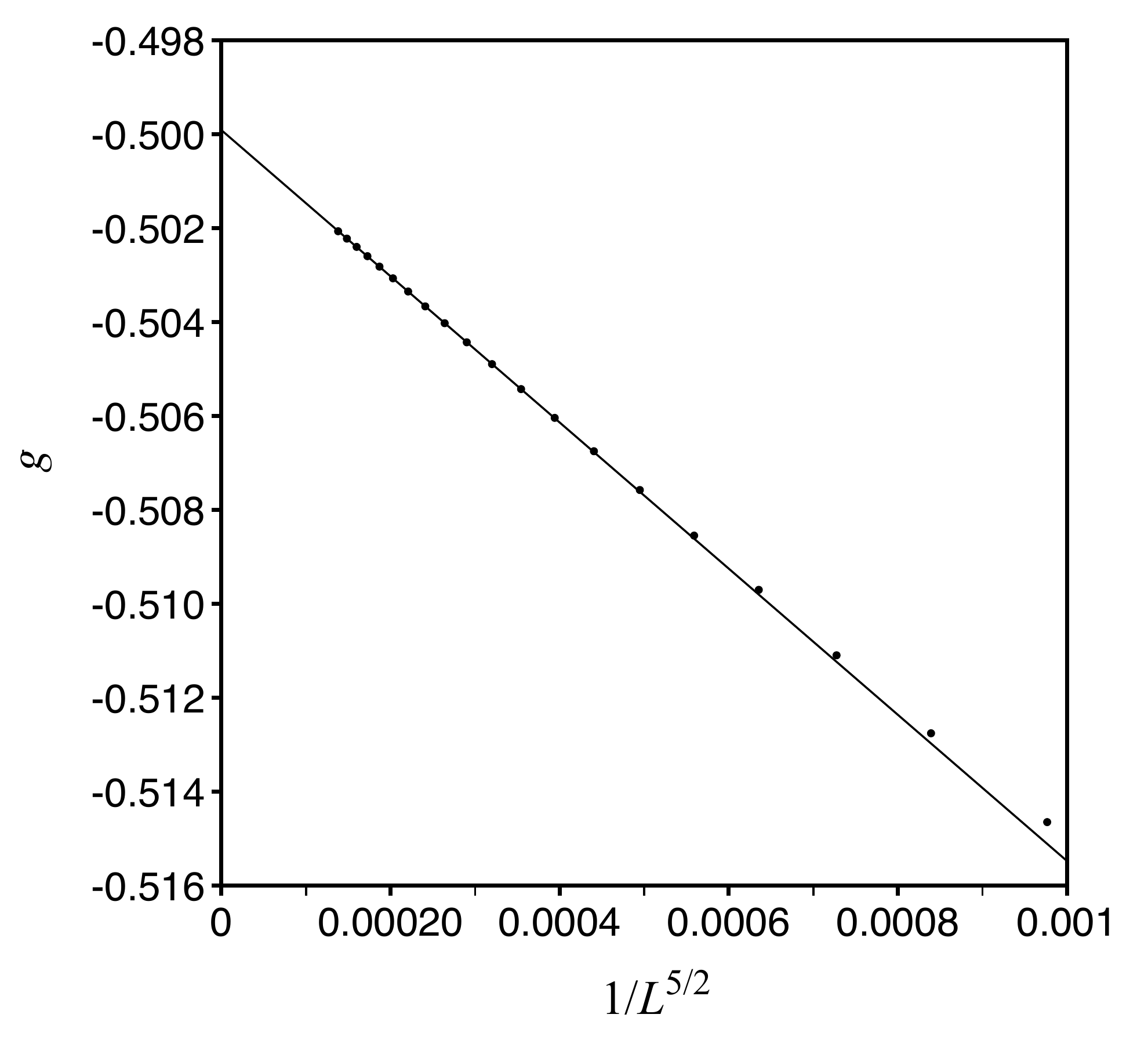}}
\end{minipage}
\hspace{0.03\textwidth}
 \begin{minipage}[t]{0.45\textwidth} 
\centerline{\includegraphics[width=\textwidth,angle=0]{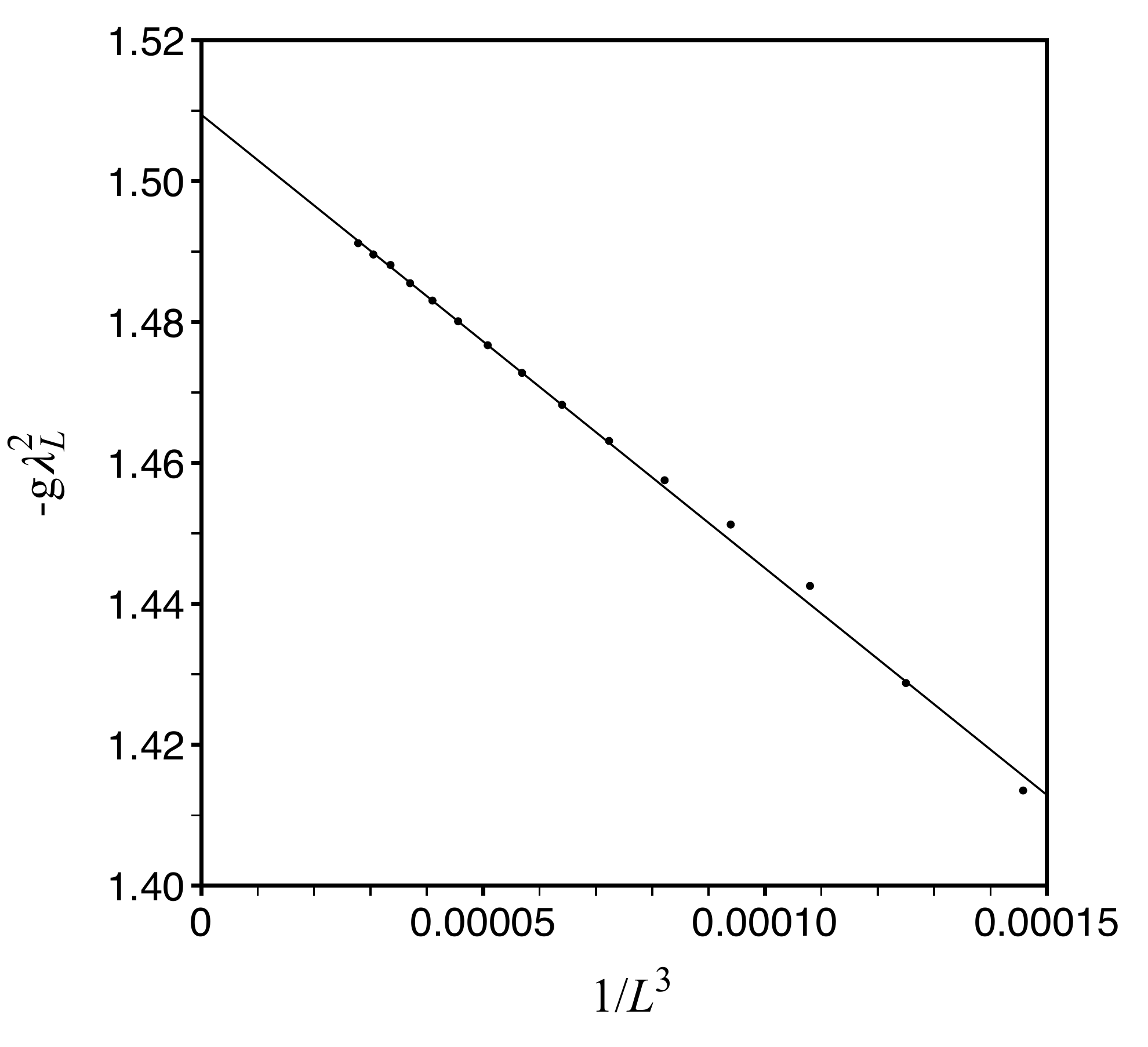}}
\end{minipage}
\caption{\label{fig:sap-para} Values of the estimators of $b \log{\lambda},$ $c \log{\lambda},$ $g$, and $-g\lambda_L^2$, for cycles in a square grid. }
\end{figure}

Finally, we fit to the assumed form by writing $$\log{d_L} \sim b\log(\lambda)  L + c \log(\lambda) + g \log{L}.$$ We then use successive triples of values $( \log{d_{k-1}},\log{d_{k}},\log{d_{k+1}}),$ with $k=2,3,\cdots, L_{\textrm{max}}-1,$ to give estimates of the parameters $b \log{\lambda},$ $c \log{\lambda},$ and $g.$ Plots of these against various powers of $1/L$ are shown in \Fref{fig:sap-para} as is a plot $-g\lambda_L^2$ obtained from the cubic fit to
the sequence $\{{\mathcal P}_L\}$. The intercept in the fourth plot is at $-g\lambda_L^2 \approx 1.59043$, which gives $g \approx -0.5006$. So all the methods gives a value of $g$  entirely consistent with an exact value of $-\frac12$. From the first two plots we estimate that $b \log(\lambda) = -0.02423 \pm 0.0001$, so $b=-0.0435 \pm 0.0002$ (the same value as for WCAS), and $c \log(\lambda) 0.9885 \pm 0.0005$,
and therefore $c=1.776 \pm 0.001$. 

In summary we have provided compelling numerical evidence that $\lambda$ and $b$ have the same values as for WCAS and that $c=1.776 \pm 0.002$ while $g=-0.500\pm 0.005$ and hence probably equals $-\frac12$.

\section{Resources}
\label{sec:resource}

The enumeration data and approximate coefficients for SAWs and cycles in a square together with
a Maple worksheet used for the asymptotic analysis can be found at our GitHub repository \url{https://github.com/IwanJensen/Self-avoiding-walks-and-polygons/tree/WCAS(H)}. This repository also contains the data used in \cite{GJ22}.

\section{Acknowledgements}
AJG wishes to thank the ARC Centre of Excellence for Mathematical and Statistical Frontiers (ACEMS) for support. 

\section*{References}



\end{document}